\documentclass[aps,prb,floatfix,twocolumn,amsmath,amssymb, superscriptaddress]{revtex4}

\usepackage{graphicx}
\usepackage{hyperref}
\usepackage[usenames]{color}

\begin{document}
\title{ Quantitative description of Josephson-like tunneling in $\nu_T=1$ quantum Hall bilayers  }

\author{Timo Hyart}
\affiliation{Max-Planck-Institut f\"{u}r Festk\"{o}rperforschung, Heisenbergstrasse 1, D-70569 Stuttgart, Germany}
\affiliation{Institute for Theoretical Physics, University of Leipzig, Vor dem Hospitaltore 1, D-04103 Leipzig, Germany}

\author{Bernd Rosenow}
\affiliation{Max-Planck-Institut f\"{u}r Festk\"{o}rperforschung, Heisenbergstrasse 1, D-70569 Stuttgart, Germany}
\affiliation{Institute for Theoretical Physics, University of Leipzig, Vor dem Hospitaltore 1, D-04103 Leipzig, Germany}
%


\begin{abstract}
At total filling factor $\nu_T=1$, interlayer phase coherence in quantum Hall bilayers can result in a tunneling anomaly resembling the Josephson effect in the presence of strong fluctuations. The most robust experimental signature of this effect is a strong enhancement of the tunneling conductance at small voltages. The height and width of the conductance peak depend strongly on the area and tunneling amplitude of the samples, applied parallel magnetic field and temperature. We find that the tunneling experiments are in quantitative agreement with a theory which treats fluctuations due to meron excitations phenomenologically and takes tunneling into account perturbatively.  We also discuss the qualitative changes caused by larger tunneling amplitudes, and provide a possible explanation for recently observed critical currents in counterflow geometry.
\end{abstract}

\maketitle

\section{Introduction}

The existence of a bilayer quantum Hall state at total filling factor $\nu_T=1$ is well-established both theoretically
\cite{Perspectives, Girvin-summer-school, Girvin-Phys-tod, Eisenstein-MacDonald,chakraborty87, fertig89, Yang94,WenandEzawa,Ezawa-Fraunhofer,Stern01,Balents01,Fogler-Wilczek01,Joglekar01,Fertig03,Fertig05,Huse05,Wang05,Rossi05,Roostaei08,Eastham09,Fil09,Sun10,Su10,Eastham10, LeeEC10, phase-transition}  and experimentally
\cite{Murphy94,Spielman00,Spielman01,quantized-Hall-Drag1,quantized-Hall-Drag2,Eisenstein03,Spielman04,Spielman-diss,Kellog04,Tutuc04,Tutuc05,Wiersma04-06,Spielman05,Finck08,Champagne08,Champagne08PRB,TiemannNJP08,TiemannPRB09,Tiemann-diss,Yoon,Giudici08,Finck10}. This state  is characterized by
remarkable electronic properties  such as counterflow superconductivity
and a Josephson-like enhancement of tunneling  between the
two layers.\cite{fertig89, Yang94,WenandEzawa, Ezawa-Fraunhofer, Perspectives}
Its
formation is controlled by the relative magnitude of intra- and interlayer Coulomb interactions, and therefore  depends critically on the ratio of interlayer separation $d$ and magnetic length $l_B=\sqrt{\hbar/eB}$. At large $d/l_B$, there is no quantum Hall effect and the bilayer system behaves qualitatively like two independent composite fermion systems. On the other hand, at small $d/l_B$, the interlayer Coulomb interaction induces an exotic quantum Hall state, which can be   described as  an exciton Bose-Einstein condensate or as a pseudospin ferromagnet.\cite{fertig89, Yang94,Perspectives}
Here, the  pseudospin  is formed from the two-valued {\it which layer} quantum degree of freedom, and  condensation occurs because the electrons can lower their interlayer exchange energy by entering
a state with uncertain layer index.\cite{Yang94}

Experimentally, the most spectacular effect arising due to the interlayer phase coherence at small $d/l_B$ is  a huge enhancement of the tunneling conductance at small voltages. In the incoherent state at large $d/l_B$, the tunneling in quantum Hall bilayers is strongly suppressed by a Coulomb gap,\cite{Coulomb-gap-experiments, Einsenstein95, Brown94, Turner} which is caused  by strong electronic correlations and the squeezing of wave functions by the magnetic field rather than by small geometric dimensions.\cite{Johansson-Kinaret} Namely,  in a strong magnetic field each layer is in a  correlated state of its own, and the tunneling event can be considered as an electron suddenly brought from one layer to the other, resulting in an excited state of the system.
Relaxation towards the ground state takes place when the energy is carried away by collective excitations, and this excitation  energy  must be provided by a sufficiently large
external voltage.\cite{Johansson-Kinaret} On the other hand, if the layers are closer together, the Coulomb gap decreases \cite{Einsenstein95} due to interlayer correlations \cite{Klironomos}, and finally at some critical value of $d/l_B$ a strong and sharp peak appears in the differential conductance \cite{Spielman00, Spielman-diss}, indicating the transition to an interlayer coherent state
  with
a quantized Hall drag and small counterflow resistances \cite{quantized-Hall-Drag1, quantized-Hall-Drag2, Kellog04, Tutuc04, Tutuc05, Wiersma04-06, Finck10}.  The coherent state was further characterized by determining   the dispersion relation of the collective Goldstone mode \cite{Spielman01} and the degree of spin polarization \cite{Spielman05, Giudici08, Finck10}. Moreover, the phase-transition from incoherent to coherent state \cite{Spielman04, Spielman05, Champagne08, Champagne08PRB, quantized-Hall-Drag2, phase-transition} was studied in some detail.

The Josephson effect with a zero bias supercurrent can be observed when there is a phase difference between two
superconductors separated by a tunnel junction.
In  a quantum Hall bilayer system, the individual layers are not superconducting by themselves,
and only the two layers combined can exhibit phase coherence.
However, in the coherent state of a clean system with tunnel coupling between the layers, a deviation of the pseudospin orientation from the minimum energy direction  is expected to give rise to a zero bias tunnel current between the layers  \cite{WenandEzawa,Ezawa-Fraunhofer}, in analogy to the Josephson current in a superconducting tunnel
junction. Experimentally, instead of a  zero-bias supercurrent,   a
Josephson-like enhancement of tunneling \cite{Spielman00, Spielman01, Eisenstein03, Spielman04, Spielman-diss, Spielman05, Finck08, Champagne08, Champagne08PRB, TiemannNJP08, TiemannPRB09, Tiemann-diss, Yoon} at small interlayer bias voltage was observed.
A significant amount of theoretical effort has been devoted  to understanding the finite height and width of the conductance peak of this Josephson-like tunneling at small voltages \cite{Stern01, Balents01, Fogler-Wilczek01, Fertig03, Fertig05, Wang05, Joglekar01, Rossi05, Su10} as well as the magnitude of the critical tunneling current \cite{Eastham10, LeeEC10, Fil09}. These approaches differ in how the bias voltage, quasiparticles and disorder are introduced into the theory.
In one type of approach \cite{Rossi05, Su10, Fil09}, the clean limit without charge disorder is considered.
Then, the critical current is given by the maximum current at which the order parameter can be static, and the tunneling conductance cannot be described in terms of condensate dynamics alone, because it also depends on the fermionic quasiparticles which transport the charge between leads and bulk.\cite{Rossi05, Su10}

Experimentally investigated samples are probably
not in the clean limit, and  experimental results indicate that disorder-induced topological defects, so-called merons, are important.\cite{Stern01, Balents01, Fogler-Wilczek01, Fertig03, Fertig05, Huse05, Roostaei08, Eastham09,Sun10, Eastham10, LeeEC10} Merons  carry a charge $\pm e/2$ and are characterized by their vorticity and the layer in which the charge resides.\cite{Yang94, Perspectives, Girvin-summer-school, Stern01} Because merons are nucleated by  charge disorder, they exist also at low temperatures,  and their dynamics is expected to give rise to  dissipation  and to strongly affect
 the interlayer tunneling.\cite{Stern01, Balents01, Fogler-Wilczek01, Fertig03, Fertig05, Huse05, Roostaei08, Eastham09,Sun10, Eastham10}
Merons can be taken into account in the description of Josephson-like tunneling by introducing a phenomenological vortex field.\cite{Stern01, Balents01, Fogler-Wilczek01, Eastham10, LeeEC10}
Both dynamic \cite{Stern01, Balents01} and static \cite{Fogler-Wilczek01, Eastham10, LeeEC10} vortex fields have been considered in the literature. Despite similar starting points, static and dynamic approaches predict  different characteristic features for the tunneling current.

Spatial fluctuations in the vortex field are accounted for by a  correlation length $\xi$, which governs the decay of the
pseudospin ferromagnetic order.
 Spatial fluctuations are caused by the formation of compressible puddles, which
are thought to arise due to local fluctuations in the density of dopants. If the local density of  merons created by charge disorder is sufficiently high,  the merons can screen the random potential and delocalize. As merons are vortices of the
order parameter field, spatial correlations in the order will decay on the typical distance between puddles, which in turn
is determined by the  setback distance to the dopant layer, such that one expects $\xi \sim 100-200$ nm \cite{Stern01, Eastham09}.

In the present manuscript,  we study the influence of  a dynamical vortex field on  interlayer tunneling deep inside the coherent phase. Dynamical fluctuations of the  order parameter are caused by the dynamics of merons in this picture. At temperatures lower than the  energy gap but still comparable to it, one can imagine that thermally activated hopping of merons from one puddle to another
is the dominant source of fluctuations. By mapping the quantum Hall bilayer to a classical two-dimensional XY model with a symmetry-breaking field and with disorder, Fertig and Straley \cite{Fertig03} have found that disorder nucleates
strings of overturned spins, which connect vortices and antivortices at their ends. At low temperatures, this state shows
glassy features and gives rise to anomalously large fluctuations
of the vortex field. The temperature dependence of
interlayer tunneling in a coherence network of puddles and ordered regions was analyzed in Ref. \onlinecite{Fertig05}.


We model dynamical vortex field  fluctuations  by introducing an exponential time decay of local pseudospin correlations
\cite{Stern01, Balents01}, which is governed by a correlation time $\tau_\varphi$. In this approach,
tunneling between the two layers can be treated perturbatively, and
Josephson oscillations in the presence of vortex field fluctuations give rise to the finite tunneling peak observed in the experiments. This effect strongly resembles the Josephson tunneling in small Josephson junctions in the presence of thermal or quantum fluctuations \cite{Steinbach, Ingold, IvanchenkoZilberman}.
We find that the current-voltage (I-V) characteristic
of interlayer tunneling is characterized by a voltage scale
%
\begin{equation}
V_0 = \hbar u /e\xi \label{V0}
\end{equation}
%
and by the scale for the maximum  tunnel current
%
\begin{equation}
I_0 \propto \frac{e}{\hbar} \frac{\xi^2 L^2}{\rho_s} \frac{\Delta_{SAS}^2}{l_B^4}  \ \ .
\end{equation}
%
Here,  $u$ is the velocity of the pseudospin
 wave mode, $\Delta_{SAS}$ is the tunnel coupling between the two layers, $L^2$  is the area of the sample and $\rho_s$ is
the pseudospin stiffness.
The functional form of the I-V characteristic is controlled by the temperature dependent decoherence rate
%
\begin{equation}
\alpha(T) =\frac{\xi}{u \tau_\varphi(T)} \label{alpha} \ \ ,
\end{equation}
%
which for instance enters into the
 zero bias conductance. For small values of $\alpha$ we obtain
%
\begin{equation}
G_0(T) \ = \ {I_0 \over V_0 } {2 \over \alpha(T)} \label{zerobiasG} \ \ .
\end{equation}
%
 While the zero bias
conductance increases when $\alpha \to 0$, the maximum tunnel current
saturates in this limit to a value $\pi I_0$.
An in-plane magnetic field suppresses the conductance peak, and we find that the
characteristic field scale for this suppression is $\Phi_0/(d \xi)$, where $\Phi_0 = h/e$ is the flux quantum.

We compare our theoretical predictions
to the experimentally observed dependencies of the tunneling current on sample area, tunneling amplitude, applied parallel magnetic field and temperature.
We find that  experimental results and theoretical predictions
are in excellent agreement. Despite the comparatively large number of parameters in our theory, we find that  all parameters can be determined uniquely from experimental data and that the parameter values obtained in this way  are in very good agreement with theoretical estimates.
We  find that the apparent temperature dependence of the tunneling conductance \cite{Spielman-diss}
$G_0 \propto  \exp(- T/T_0)$ can be explained as a crossover between   an activated and a power law temperature dependence of
$\alpha(T)$, which in turn
has a natural explanation  in terms of vortex field fluctuations.

More recently  \cite{TiemannNJP08, TiemannPRB09, Tiemann-diss, Yoon}, intriguing  observations were made in bilayer samples with  larger tunneling amplitude. In particular, jumps of the tunnel current were observed and described in terms of critical tunnel currents about two orders of magnitude larger than in previous experiments. We find that these observations can still be explained in the framework of a perturbative approach  \cite{Stern01, Balents01} when taking into account that the tunneling resistance between the layers is so small in these samples that the other resistances in the system, such as circuit, contact, longitudinal and Hall resistances, become important.  Our approach also provides
an explanation for recently observed critical currents in a
counterflow geometry \cite{Yoon}. This surprisingly good agreement with the experiments over a wide range of tunneling amplitudes raises questions regarding the region of validity
of the perturbative treatment of the tunneling. We give a short discussion of this important theoretical issue in Section \ref{conclusions}.

Our paper is organized as follows. In Section \ref{theory} we describe the theory of interlayer tunneling in the presence of fluctuations caused by the meron excitations. In Section \ref{Caltech} we compare the experimental tunneling I-V characteristic
obtained using samples with extremely small tunneling amplitude to our theoretical predictions. We find a good quantitative agreement between  theory and experiments. In Section \ref{largetun} we apply the theory to describe the qualitative changes observed in recent experiments where the tunneling amplitude of the samples was significantly larger. In Section \ref{others}, we compare
with previous theoretical approaches, and
finally in Section \ref{conclusions} we give a summary of our findings and discuss the limits of the validity of our approach.

\section{Theory of interlayer tunneling in the presence of meron excitations \label{theory}}

We concentrate on the tunneling deep inside the coherent phase at $\nu_T=n_0 h/(e B)=1$, where $n_0=(2 \pi l_B^2)^{-1}$ is the average density. We assume that the real spin is fully polarized in agreement with the experiments \cite{Spielman05, Giudici08, Finck10}. The pseudospin is therefore
the most relevant degree of freedom in our problem: The low-energy excitations of the system are the pseudospin waves and the merons. The theoretical analysis presented in this section largely follows that of Stern et al.~in Ref.~\onlinecite{Stern01}.  In the absence of tunneling, the pseudospin waves are described by a Hamiltonian density \cite{Yang94,Perspectives,Stern01}
\begin{equation}
 \mathcal{H}=\frac{1}{2} \rho_s (\nabla \varphi)^2+\frac{(e n_0 m_z/2)^2}{2 \Gamma}, \label{startingpoint}
\end{equation}
where $\vec{m}=(\cos \varphi, \sin \varphi, m_z)$ is the pseudospin vector, $\rho_s$ is the pseudospin stiffness and $\Gamma$ is the capacitance per area. The momentum conjugate to
$\varphi$ is $\pi=\hbar n_0 m_z/2$. Physically the first term in equation (\ref{startingpoint}) arises from the loss of optimal Coulomb exchange energy: If the pseudospin varies in space, the spatial part of the many-particle wave function cannot be fully antisymmetric. The second term measures the capacitive energy. Because of exchange effects between the layers  $\Gamma$ is expected to be strongly enhanced from its electrostatic value.\cite{Yang94, Stern01, Fil09} The dispersion relation for collective pseudospin waves (Goldstone mode) is
\begin{equation}\omega_{\vec{k}}=u k, \label{dispersion}
\end{equation}
where the velocity of the pseudospin waves $u$ is
\begin{equation}
u=\sqrt{\frac{\rho_s}{\Gamma}} \frac{e}{\hbar}. \label{velocity}
\end{equation}

The tunneling energy should be included in the Hamiltonian (\ref{startingpoint}) as $T_++T_-$ \cite{Yang94,Perspectives,Stern01}, where the tunneling operators are
\begin{equation}
 T_{\pm}=- \lambda \int d^2 r \ e^{\pm i \varphi} e^{\pm i \varphi_m} e^{\pm i eVt/\hbar} e^{\pm iQ_B x}. \label{tunnelingops}
\end{equation}
Here $\lambda=\Delta_{SAS}/(8 \pi l_B^2)$, $Q_B=eB_{||}d/\hbar$, $B_{||}$ is the in-plane magnetic field and $V$ is the interlayer voltage, which we assume to be constant in space. The merons are included phenomenologically with the help of the vortex field $\varphi_m$.\cite{Stern01, Balents01, Fogler-Wilczek01, Eastham10} Using the commutation relation for $\varphi$ and $\pi$ it is easy to verify that the tunneling operators $T_\pm$ change the charge on the capacitor plates by $\pm e$.

Both counterflow and tunneling experiments suggest that at least part of the merons are mobile. Therefore,
we assume that the vortex field is dynamic. The tunneling current depends on the details of vortex field dynamics, which we assume to be characterized by a correlation length $\xi$ and a correlation time $\tau_\varphi$. For the correlation function we make the specific assumption
%
\begin{equation}
\langle e^{i \varphi_m(\vec{r}, t)} e^{-i \varphi_m(\vec{0}, 0)} \rangle =e^{-r/\xi}e^{-|t|/\tau_\varphi}\ \ ,
\label{disordercor}
\end{equation}
%
which appears to produce a good agreement between theory and experimental results.  By studying the experimentally found dependence of the tunneling current on a parallel magnetic field, we find strong evidence that $\xi \sim 100-200$ nm, in agreement with  earlier theoretical estimates \cite{Stern01, Eastham09}. Because pseudospin fluctuations are uncorrelated on scales much larger than $\xi$,  we can at least qualitatively think of the small domains of size $\xi^2$ as individual Josephson junctions. These small domains are well-coupled to each other by counterflow and transport currents. We assume that the tunneling can be considered perturbatively due to the reasonably small correlation time $\tau_\varphi$ caused by the strong fluctuations of the vortex field \cite{Stern01}. The validity of the perturbative treatment of the tunneling has been well-established \cite{Ingold} in small Josephson junctions in the presence of fluctuations, but the justification of similar assumption in quantum Hall bilayers is questionable especially at low voltages \cite{Balents01, Fogler-Wilczek01} essentially because the phase $\varphi$ describes both the counterflow currents and the tunneling simultaneously. Nevertheless, we find that the perturbative treatment of the tunneling gives a good quantitative description of the Josephson-like tunneling for all voltages and a surprisingly large range of tunneling amplitudes. We discuss this unexpected result and the limits of validity of our approach in more detail in Section \ref{conclusions}.  In this section we concentrate on a situation where the tunneling resistance is larger than the intralayer resistances, $R_{xx}$ and $R_{xy}$, so that the interlayer voltage can be considered homogeneous. However, we stress that our approach can easily be extended to a case where the interlayer voltage varies slowly in space. We believe that the interlayer voltage is very homogeneous in the experiments where the tunneling amplitude is small \cite{Spielman00, Spielman01, Eisenstein03, Spielman04, Spielman-diss, Spielman05, Finck08, Champagne08, Champagne08PRB}. In the case of larger tunneling amplitudes the inhomogeneities will affect the details of the tunneling I-V characteristics, but we believe that they still do not play an important role in the main qualitative features observed in the experiments reported in Refs.~\onlinecite{TiemannNJP08, TiemannPRB09, Tiemann-diss, Yoon} (see Section \ref{largetun}).

The I-V characteristics can now be calculated using Eqs.~(\ref{startingpoint})-(\ref{tunnelingops}) and  Fermi's golden rule, similar to the case of Josephson junctions \cite{Ingold, Ingold-Nazarov}. The result of this calculation is
\begin{eqnarray}
I(V, B_{||})&=&\frac{4 e \lambda^2 L^2}{\hbar^2}\int_0^\infty dt \int d^2 r \ G_m(r,t) e^{-D(r,t)/2} \nonumber\\
&&\times \sin \frac{C(r,t)}{2} \cos(Q_B x) \sin \frac{eVt}{\hbar}, \label{main1}
\end{eqnarray}
where
\begin{equation}
 D(r,t)=\sum_{\vec{k}} \frac{\hbar u}{L^2 \rho_s k}[1-\cos(\vec{k}\cdot\vec{r}) \cos(ukt)] \coth\frac{\beta \hbar u k}{2}, \label{main2}
\end{equation}
\begin{eqnarray}
 C(r,t)&=&\sum_{\vec{k}} \frac{\hbar u}{L^2 \rho_s k}\sin(uk t) \cos(\vec{k}\cdot \vec{r})\nonumber\\&=&\frac{\hbar}{2 \pi \rho_s} \theta(ut-r)\bigg[t^2-\frac{r^2}{u^2}\bigg]^{-1/2}, \label{main3}
\end{eqnarray}
$\beta=1/k_BT$ and $G_m(r,t)=\langle e^{i \varphi_m(\vec{r}, t)} e^{-i \varphi_m(\vec{0}, 0)} \rangle$ is the correlation function for $\varphi_m$.  This result has been found previously in Ref.~\onlinecite{Stern01}, but for completeness we give a detailed derivation of equations (\ref{main1})-(\ref{main3}) in Appendix \ref{calculation1}.


In order to calculate $D$, we need to define an ``ultraviolet cut-off momentum'' \cite{Stern01} $k_0 = \kappa \sqrt{2}/l_B$, where it is expected that $\kappa \approx 1$. Although some of the parameters, like $\rho_s$, $\Gamma$ and $\Delta_{SAS}$, can be theoretically estimated \cite{Yang94, Stern01, Fil09}, there remain reasonably large uncertainties in their actual values. We find that it is possible to determine most of the parameters based on experimental data and the obtained values are in good agreement with the theoretical predictions. Before comparing  theory with experiments, we will make certain simplifying assumptions. The theoretically estimated value for pseudospin stiffness is $\rho_s \sim 0.4$ K \cite{Stern01}, and according to  Ref.~\onlinecite{Su10} this parameter is likely to be the most reliably known continuum model parameter. Using this value of $\rho_s$ and the values of the other parameters found below, we have numerically checked that $D$ depends only weakly on temperature in the experimentally relevant temperature range. Therefore, we can use everywhere the zero temperature value
\begin{equation}
D_0 \approx \frac{\hbar u k_0}{2\pi \rho_s} = \kappa \sqrt{\frac{e^3 B}{2 \pi^2 \rho_s \Gamma \hbar}}.
\end{equation}
For $B = 2$ T, we obtain $D_0 \sim 3.3$, but an accurate estimate is not possible due to the uncertainty related to the ultraviolet cut-off momentum. With similar assumptions, we have numerically checked that we can use the following first order expansion in Eq.~(\ref{main1})
\begin{equation}
 \sin \frac{C(r,t)}{2} \approx \frac{C(r,t)}{2}.
\end{equation}
These simplifications were justified also in Ref.~\onlinecite{Stern01}.

By using the Eqs.~(\ref{main1})-(\ref{disordercor}) and the approximations outlined
above, we get ($R=r/\xi$ and $q=k\xi$)
\begin{eqnarray}
I&=& I_0  \int d q   \bigg[\frac{\alpha}{\alpha^2+(V/V_0-q)^2} -\frac{\alpha}{\alpha^2+(V/V_0+q)^2} \bigg] \nonumber\\&& \hspace{0.5 cm} \times \int dR \ R \ e^{-R} J_0(q R) J_0(Q_B \xi R),
\label{altapp}
\end{eqnarray}
where
\begin{equation}
I_0=\frac{e}{\hbar} \frac{\xi^2 L^2}{\rho_s} \frac{\Delta_{SAS}^2}{l_B^4} \frac{e^{-D_0/2}}{64 \pi^2}, \label{I0exp}
\end{equation}
and $V_0$ and $\alpha$ are given by Eqs.~(\ref{V0}) and (\ref{alpha}), respectively.
In Appendix \ref{calculation2} we give an alternative derivation of Eq.~(\ref{altapp}) where the quantum fluctuations are neglected. Similarly as in Ref.~\onlinecite{Fogler-Wilczek01}, we solve the Sine-Gordon equation perturbatively, but in contrast to Ref.~\onlinecite{Fogler-Wilczek01} we also take into account the dynamics of the vortex field. Finally, we obtain the same expression~(\ref{altapp}) for the I-V characteristics except that now the factor $e^{-D_0/2}$ does not appear in the equation for $I_0$. We interpret this factor as the effect caused by quantum fluctuations.

In the absence of a magnetic field, $Q_B=0$, we obtain
\begin{eqnarray}
I(V)
&=& I_0  \int d q   \bigg[\frac{\alpha}{\alpha^2+(V/V_0-q)^2}\nonumber\\&& \hspace{0.5 cm}-\frac{\alpha}{\alpha^2+(V/V_0+q)^2} \bigg] \frac{1}{(1+q^2)^{3/2}}.
\label{altapp2}
\end{eqnarray}
In the limit $V \to 0$ and for small values of $\alpha$, the peak conductance $G_0 = \left. {d I \over
d V}\right|_{V=0}$ is given by Eq.~(\ref{zerobiasG}). The maximum tunneling current in the limit $\alpha \to 0$ saturates to a value $\pi I_0$.

We expect that $V_0$, $I_0$ and $\xi$ depend only weakly on temperature. The strong temperature dependence observed in experiments would then be caused by the temperature dependence of $\tau_\varphi$. In our opinion, this assumption is in agreement with  more detailed theoretical models where the meron dynamics were analyzed \cite{Fertig03, Fertig05, Roostaei08}.  We expect that for temperatures comparable to the thermal activation gap $\Delta_g$, the temperature dependence of $\tau_\varphi$ arises mainly due to thermally activated hopping of the merons. On the other hand, the temperature dependence of $\tau_\varphi$ at temperatures much smaller than $\Delta_g$ could originate from glassy excitations of the pinned meron system, which were predicted \cite{Fertig03} to act as a dissipative environment.

The determination of the different theoretical parameters of the model is discussed in detail in Section \ref{Caltech}. We summarize the methods used to determine them and the values  we obtain in table \ref{taulukko}. The values of the parameters obtained using the experimental data are in good agreement with the theoretical expectations \cite{Yang94, Stern01, Fil09, Eastham09}.

\begin{table}[t]
\centering
\begin{tabular}{|c|c|c|}
	\hline
Parameter &  Method of determination   & Value  \\
	\hline
    \hline
$\rho_s$  & Theoretical estimate &  $\sim 0.4$ K \\
    \hline
$\xi$ & Dependence of the tunneling  & $\sim 130$ nm \\
      & conductance on $B_{||}$       & \\
\hline
$u$ & Resonant enhancement of the  & $\sim 14$ km/s \\
    &   current at $eV=\hbar u Q_B$ & \\
\hline
$\Gamma$ & Using estimates for $\rho_s$ and $u$ & $\sim 10 \epsilon \epsilon_0/d$ \\
\hline
$\tau_\varphi$ & Height and width of the & Temperature \\
             &conductance peak           & dependent  \\
\hline
$\Delta_{SAS}$ &Tunneling I-V characteristics  & $\sim 10$ $\mu$K\\
 &             at $\mathbf{B}=0$ & $\sim 100$ $\mu$K \\
 \hline
\end{tabular}
\caption{Determination of the parameters. The different estimates for $\Delta_{SAS} \sim 10$ $\mu$K and $\Delta_{SAS} \sim 100$ $\mu$K correspond to samples considered in Sections \ref{Caltech} and \ref{largetun}, respectively.}
\label{taulukko}
\end{table}

\section{Quantitative description of Josephson-like tunneling in the case of small $\Delta_{SAS}$ \label{Caltech}}

We start by considering the first experimental observations of the Josephson-like tunneling \cite{Spielman00, Spielman01, Eisenstein03, Spielman-diss}. Our theory should be most applicable to these experiments, because the tunneling amplitudes of the samples were extremely small, and therefore  higher order tunneling processes can be safely neglected. In addition, the tunneling resistance is larger than the intralayer resistances $R_{xx}$ and $R_{xy}$, so that the interlayer voltage can be considered to be constant in space. We expect that in  samples with sufficiently small $\Delta_{SAS}$  tunneling is a bulk phenomena which takes place homogeneously throughout the whole sample area, independent of the  sample geometry.
Recently, it was found \cite{Finck08} that the peak in the tunneling conductance is proportional to the area of the sample. Moreover, it seems that in these experiments the tunneling current at all voltages scales proportionally to the area. These observations are explained naturally within the present approach, which
predicts the current scale $I_0$ to be proportional to the area [see Eq.~(\ref{I0exp})].

The layer separation and the surface area in these experiments \cite{Spielman00, Spielman01, Eisenstein03, Spielman-diss} are $d=27.9$ nm and $L^2=250 \cdot 250$ $\mu$m$^2$, respectively. The tunneling amplitude can be determined \cite{Spielman-diss} by solving the Schr\"{o}dinger equation for the specific structure. Another possibility is to use the reasonably well-understood \cite{Turner, Murphy95} I-V  characteristics at $B=0$ as explained in Ref.~\onlinecite{Spielman-diss}. These two methods are in good agreement with each other \cite{Spielman-diss}, and therefore we are reasonably confident that $\Delta_{SAS}$ in these experiments is approximately $5-10$ $\mu$K. We discuss the experimental and theoretical estimates for the other parameters $\xi$, $\tau_\varphi$ and $u$ below.

\subsection{Dependence on parallel magnetic field}

\begin{figure}
\includegraphics[scale=0.5]{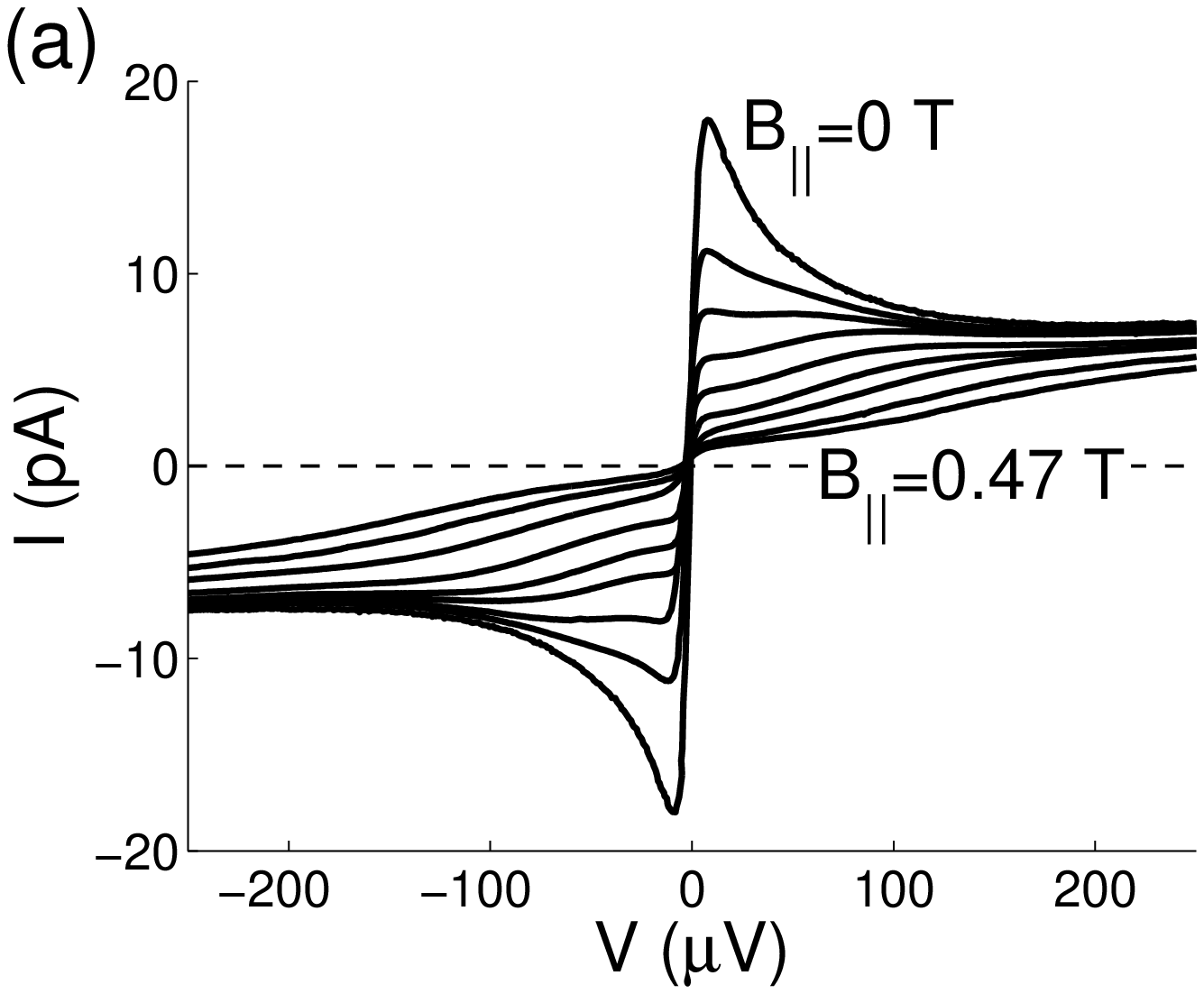} \includegraphics[scale=0.5]{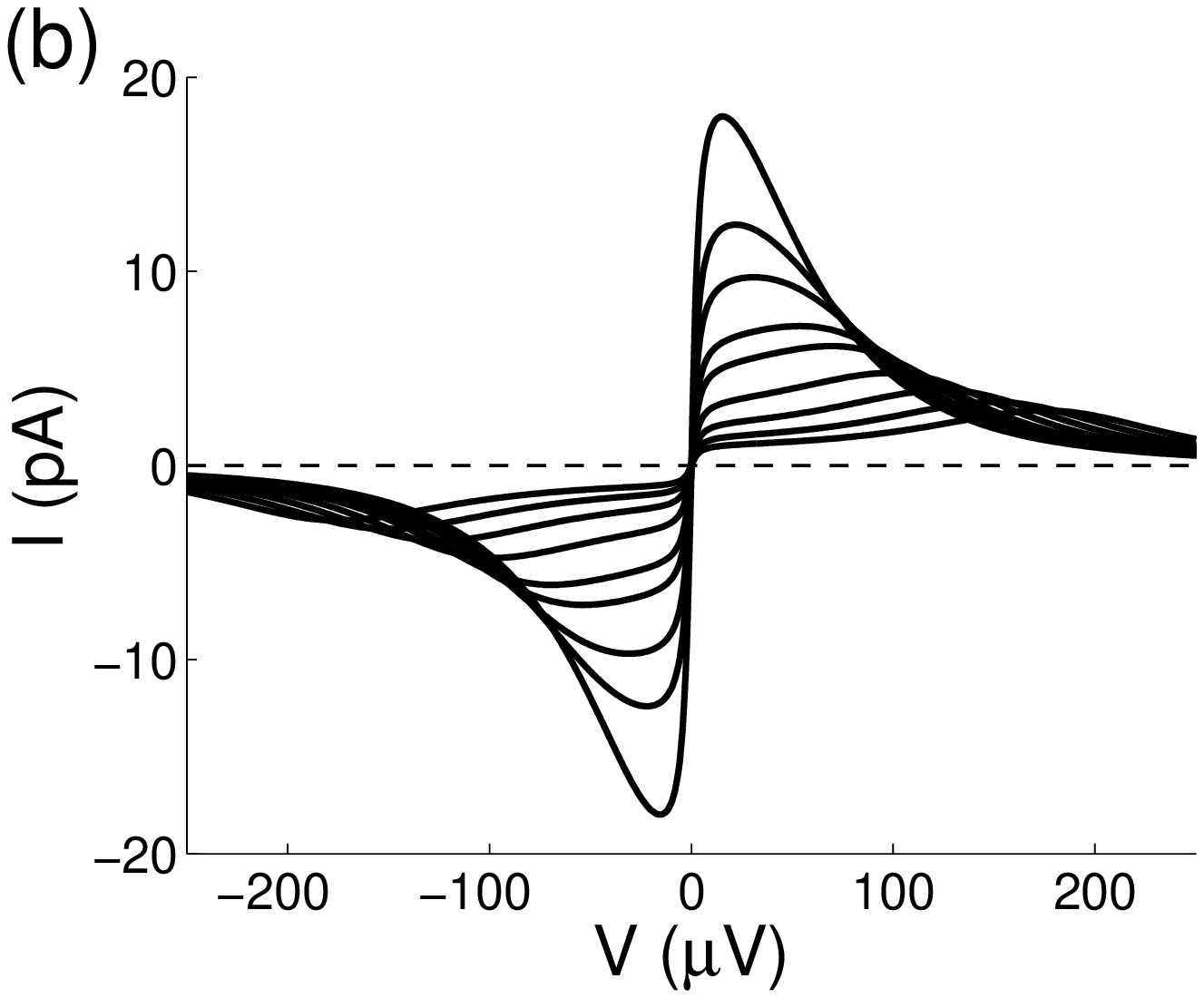}
\caption{(a) Measured I-V characteristics for $d/l_B=1.61$, $T=25$ mK and different values of parallel magnetic field $B_{||}=:0,0.11, 0.15, 0.20, 0.23, 0.29, 0.35, 0.41, 0.47$ T. (b) Theoretical I-V characteristics obtained using Eq.~(\ref{altapp}) for $V_0=73$ $\mu$V, $I_0=7.1$ pA, $\alpha=0.04$,  $\xi=126$ nm and the same values of parallel magnetic field. The determination of the theoretical parameters is described in the text. The experimental curves are reproduced from the experimental data provided by I. B. Spielman \textit{et al.} and reported in Ref.~\onlinecite{Spielman-diss}.}
\label{spielman63}
\end{figure}

The in-plane magnetic field in quantum Hall bilayers has two main effects on the Josephson-like tunneling. It suppresses the small-bias current and results in resonant enhancements of the tunneling current at voltages satisfying $eV=\hbar u Q_B$ (see Fig.~\ref{spielman63}).  In this section we discuss the physics behind these effects and determine the values of $\xi$ and $u$ by comparing the theory and experiments. We concentrate on the low voltage part of the I-V characteristics where the experimentally measured [Fig.~\ref{spielman63} (a)] and theoretically calculated [Fig.~\ref{spielman63} (b)] tunneling currents are in good agreement with each other. The theoretically calculated tunneling currents at large voltages are typically significantly smaller than the experimentally measured currents [cf. Figs.~\ref{spielman63} (a) and (b)]. The origin of this discrepancy is probably an incoherent contribution to tunneling discussed in more detail in Section \ref{conclusions}.

\begin{figure}
\includegraphics[scale=0.52]{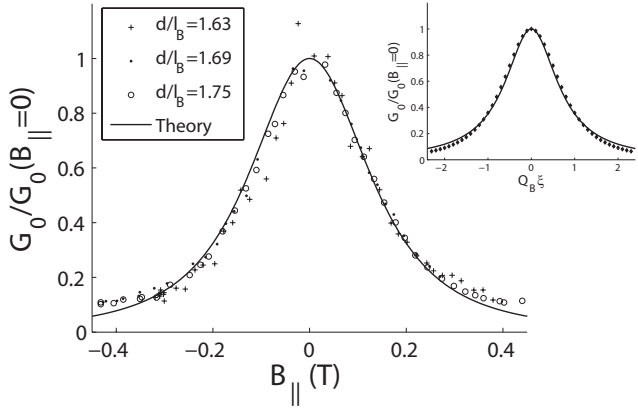}
\caption{Experimentally measured and theoretically calculated small bias tunneling conductance as a function of parallel magnetic field $B_{||}$. The experimental results were obtained at $T=25$ mK.  The small-bias tunneling conductance was calculated using Eq.~(\ref{altapp}) for $\alpha=0.04$ and $\xi = 126$ nm. The value of $\xi$ was obtained by fitting Lorentzians to the theoretical and experimental results and by setting the widths of the Lorentzians equal to each other. Inset shows the theoretical small bias conductance as a function of $Q_B \xi$ (dots) and the corresponding Lorentzian fit (line). The experimental data was provided by I. B. Spielman \textit{et al.} and has been reported in Ref.~\onlinecite{Spielman-diss}.}
\label{spielman66}
\end{figure}

As already pointed out the domains of size $\sim \xi^2$ can, at least qualitatively, be considered as individual Josephson junctions, which are well-coupled to each other by counterflow and transport currents. If a magnetic flux $\Phi$ is applied across a Josephson junction, it tries to induce a circulating current inside the junction.\cite{Tinkham} Due to this effect, the critical tunneling current of the Josephson junction oscillates as a function of $\Phi/\Phi_0$, where $\Phi_0$ is the magnetic flux quantum. The current completely vanishes if an integer number of flux quanta is applied across the junction and the dependence, in general, is typically referred to
as Fraunhofer diffraction pattern by the analogy with the case of light passing through a narrow rectangular slit.\cite{Tinkham} Based on our analogy, we can now define for each domain a magnetic flux penetrating the area between the layers as $\Phi=B_{||}\xi d$. Similarly
to the case of Josephson junctions, we expect that the current vanishes quickly as a function of $\Phi/\Phi_0$.  The important difference to standard Josephson junctions is that in our case, the size of the junctions is not well-defined and therefore the Fraunhofer diffraction pattern is "washed away". Here the exact form of the $r$ dependence of the vortex-field correlation function $G_m(r, t)$ also becomes important. It defines the size distribution of our Josephson junctions and therefore it strongly affects the shape of the small-bias tunneling conductance plotted as a function of $B_{||}$. We find that the exponential dependence given by Eq.~(\ref{disordercor}) gives significantly better agreement with the experiments than the Gaussian dependence suggested in Ref.~\onlinecite{Stern01}.

Figure \ref{spielman66} shows the experimentally measured small bias tunneling conductance as a function of $B_{||}$ and the corresponding theoretical result obtained using Eq.~(\ref{altapp}). Both dependencies are approximately Lorentzian as demonstrated for the theoretical curve in the inset of Fig.~\ref{spielman66}.
The half width at half maximum (HWHM) of the Lorentzians fitted to the experimental and theoretical curves are $B_{||}^{\textrm{HWHM}} \approx 0.138$ and $Q_B^{\textrm{HWHM}} \xi \approx 0.735$.
Using the relation $Q_B = e B_{||} d/\hbar$, we find $\xi \approx 126$ nm. This method of determining $\xi$ is extremely robust against variation of the other theoretical parameters. The obtained value is completely independent of the values of $I_0$ and $V_0$, and it changes only by a few percent if the parameter $\alpha$ is varied between $0.001$ and $0.1$. Moreover, the value of $\xi$ obtained in this way is in excellent agreement with the theoretical expectations in Refs.~\onlinecite{Stern01, Eastham09, Eastham10}.
It is also important to note that in clean samples a Fraunhofer pattern has been predicted \cite{Ezawa-Fraunhofer}, and much smaller magnetic fields will induce
circulating tunneling currents. Therefore, the in-plane magnetic field dependence of the small-bias tunneling conductance clearly manifests the important role of the merons in the Josephson-like tunneling.

\begin{figure}
\includegraphics[scale=0.5]{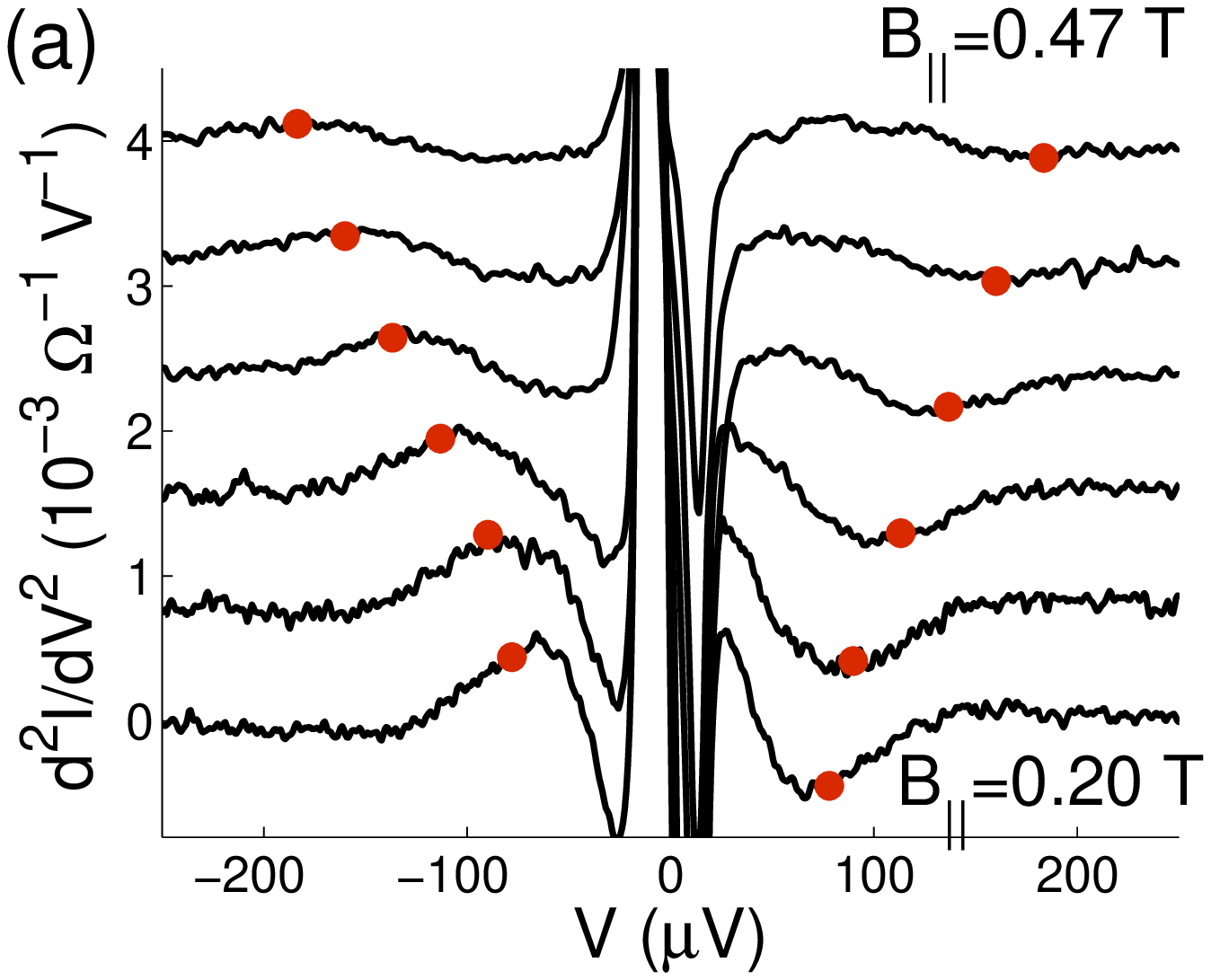} \includegraphics[scale=0.5]{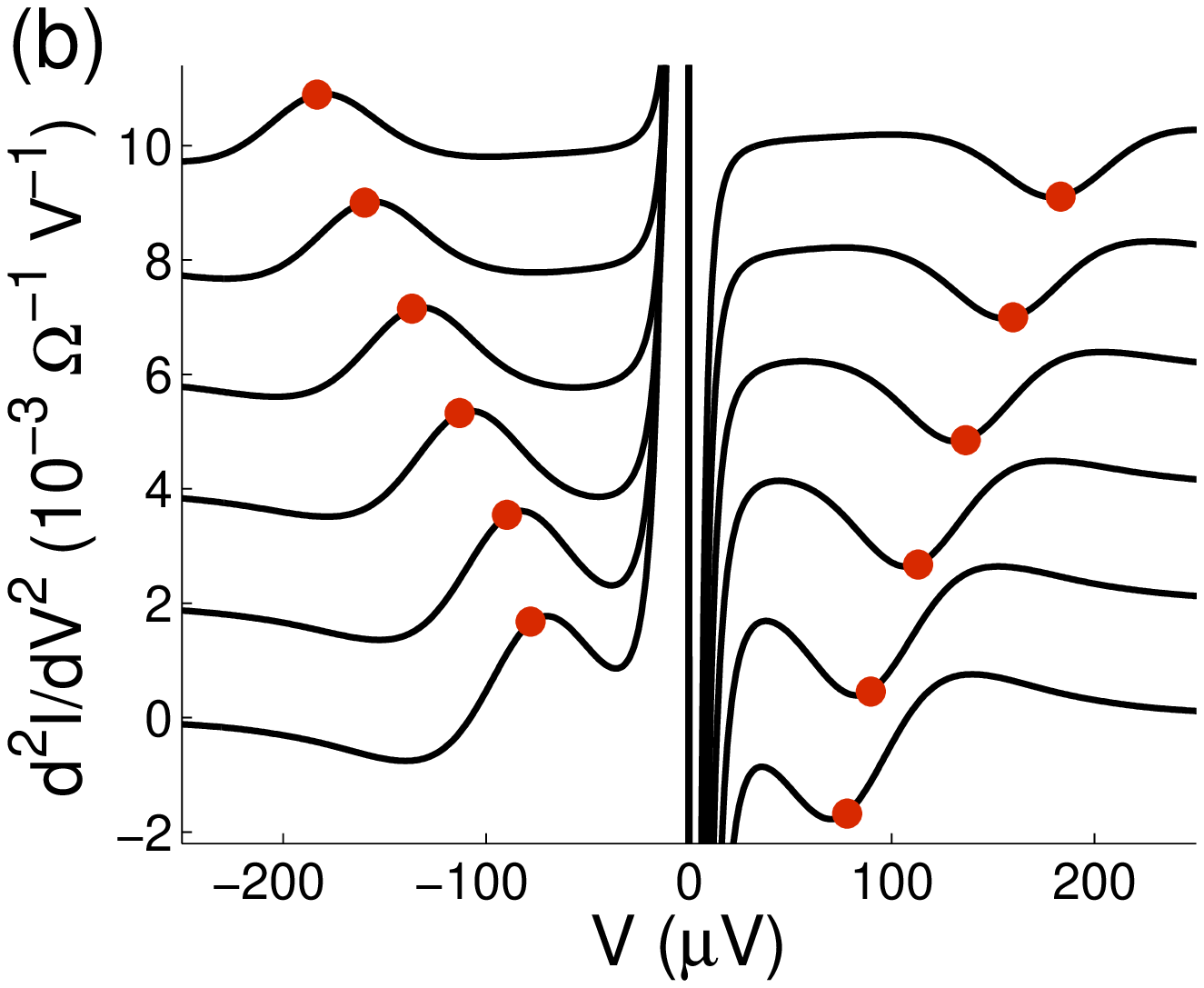}
\caption{(color online).   (a) Experimental measurement of $d^2I/dV^2$ for different values of parallel magnetic field $B_{||}=:0.20, 0.23, 0.29, 0.35, 0.41, 0.47$ T. (b) Theoretical curves obtained using $V_0=73$ $\mu$V, $I_0=7.1$ pA, $\alpha=0.04$, $\xi=126$ nm and same values of $B_{||}$. Dots show the expected locations of the resonances $V^*=\hbar uQ_B/e$, which for sufficiently large $Q_B$ are very close to the local extrema of $d^2I/dV^2$. In theoretical curves we have introduced larger offset ($2 \cdot 10^{-3}$ $\Omega^{-1} V^{-1}$) between the curves than in experimental results ($0.8 \cdot 10^{-3}$ $\Omega^{-1} V^{-1}$). The experimental curves are reproduced from the experimental data provided by I. B. Spielman \textit{et al.} and reported in Ref.~\onlinecite{Spielman-diss}.}
\label{spielman65}
\end{figure}

The second interesting effect caused by the in-plane magnetic field are the additional enhancements of the tunneling current at voltages $V^*(B_{||})$ satisfying $eV^*=\hbar uQ_B$.  The origin of these resonances is nicely explained in Refs.~\onlinecite{Stern01, Balents01, Fogler-Wilczek01}. The parallel magnetic field allows only tunneling between states that differ by a momentum $Q_B$, and energy conservation requires that the energy of these states differs by $eV$. In the absence of a dissipative environment, there would be just a single linearly dispersing collective mode $\omega_{\vec{k}}=uk$ and therefore both conditions can be satisfied only if the voltage satisfies the resonance condition $eV^*=\hbar u Q_B$. Therefore the resonant enhancements of the tunneling current in the I-V characteristic essentially map the dispersion relation of the pseudospin waves.\cite{Stern01, Balents01, Fogler-Wilczek01} Because of the finite correlation length and time caused by the merons,  these resonances are not sharp but they can still be clearly seen both in the experimental and theoretical curves by plotting the second derivative $d^2I/dV^2$ as a function of $V$ as shown in Fig.~\ref{spielman65}. The second derivative is not sensitive to any linear background conductance and therefore it shows more clearly the nonlinear features in the I-V characteristics. In particular the voltages $V^*$ at the resonances can be determined by finding the extrema of $d^2 I/dV^2$, which indicate a maximum of the curvature of the I-V characteristic. This result was already anticipated in Refs.~\onlinecite{Spielman01} and \onlinecite{Spielman-diss}, where the velocity $u$ of the pseudospin waves was obtained experimentally by determining for each magnetic field $B_{||}$ the voltage $V^*$ where  $d^2 I/dV^2$ has an extremum, and by fitting a linear relation
$V^* = \hbar u Q_B/e$ to the data points. We have theoretically confirmed the validity of this approach by numerically determining the locations $V^*$ of extrema of $d^2 I/dV^2$ for different values of $Q_B$. For $Q_B \xi
\gtrsim 2$, the numerically determined function $V^*(Q_B)$ can be approximated by the linear relation $V^* = \hbar u Q_B/e$ within an accuracy of few percents. In addition, the numerically determined function  $V^*(Q_B)$
is completely independent of $I_0$ and practically independent of $\alpha$, i.e.~it does not change within our numerical accuracy  when $\alpha$ is varied between $0.001$ and $0.1$.
Therefore, the experimentally found velocity $u \approx 14$ km/s can be considered as a reliable estimate of the actual velocity of  pseudospin waves.  This value of $u$ is somewhat smaller than theoretically expected \cite{Yang94, Spielman01, Spielman-diss, Stern01, Fil09}, but probably still within the theoretical uncertainties because the parameters $\rho_s$ and $\Gamma$ are not known accurately.

By using our estimates for the correlation length $\xi$ and Goldstone mode velocity $u$, we find $V_0 \approx 73$ $\mu$V, which yields good agreement between theoretical and experimental curves  in Figs.~\ref{spielman63} and \ref{spielman65}. We can now determine the parameters $\alpha$ and $I_0$ by using the measured values of small-bias tunneling conductance $G_0=4.68 \cdot 10^{-6} \ \Omega^{-1}$ and maximum tunneling current $I_{\max}=18$ pA at $B_{||}=0$. We obtain $\alpha \approx 0.04$ and $I_0 \approx 7.1$ pA.
By assuming that $\rho_s \approx 0.4$ K, we estimate using Eq.~(\ref{I0exp}) that $I_0 \sim 4-16$ pA (corresponding to $\Delta_{SAS}=5-10$ $\mu$K and $d/l_B=1.6$). Therefore the values of the parameters $V_0$, $I_0$ and $\xi$ determined by fitting the theory to the experimental data shown in Figs.~\ref{spielman63}, \ref{spielman66} and \ref{spielman65} are consistent with Eq.~(\ref{I0exp}). Moreover, we find that the exchange enhanced capacitance $\Gamma$ is more than $10$ times larger than the electrostatic value. This means that an interlayer bias $V \sim 100$ $\mu$V can cause a density imbalance of $5 \%$. We believe that the density imbalance induced by the interlayer voltage plays an important role in the recent tunneling experiments performed in the counterflow geometry (see Section \ref{largetun}).

\subsection{Temperature dependence \label{sec:tempdep}}

\begin{figure}
\includegraphics[scale=0.5]{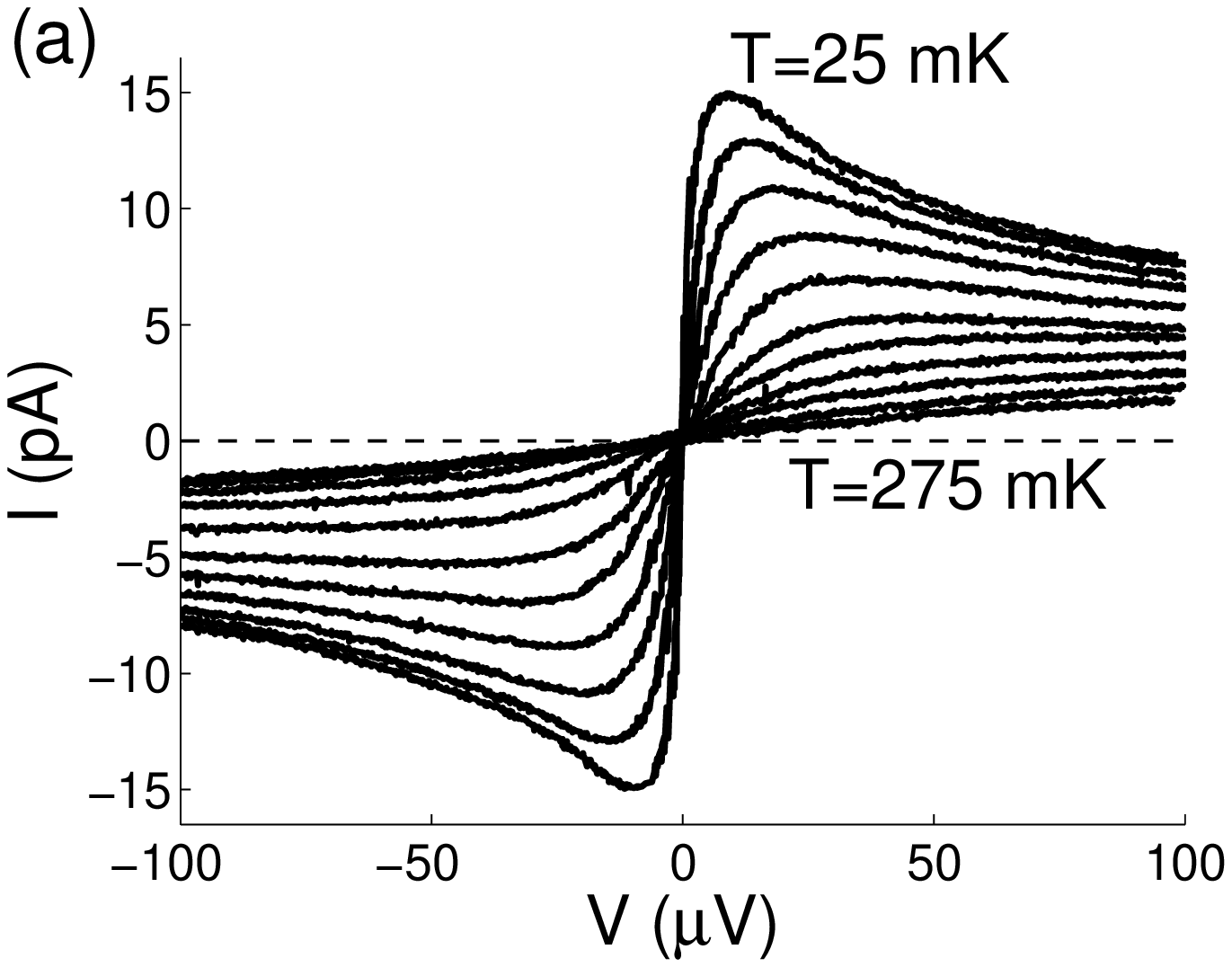} \includegraphics[scale=0.5]{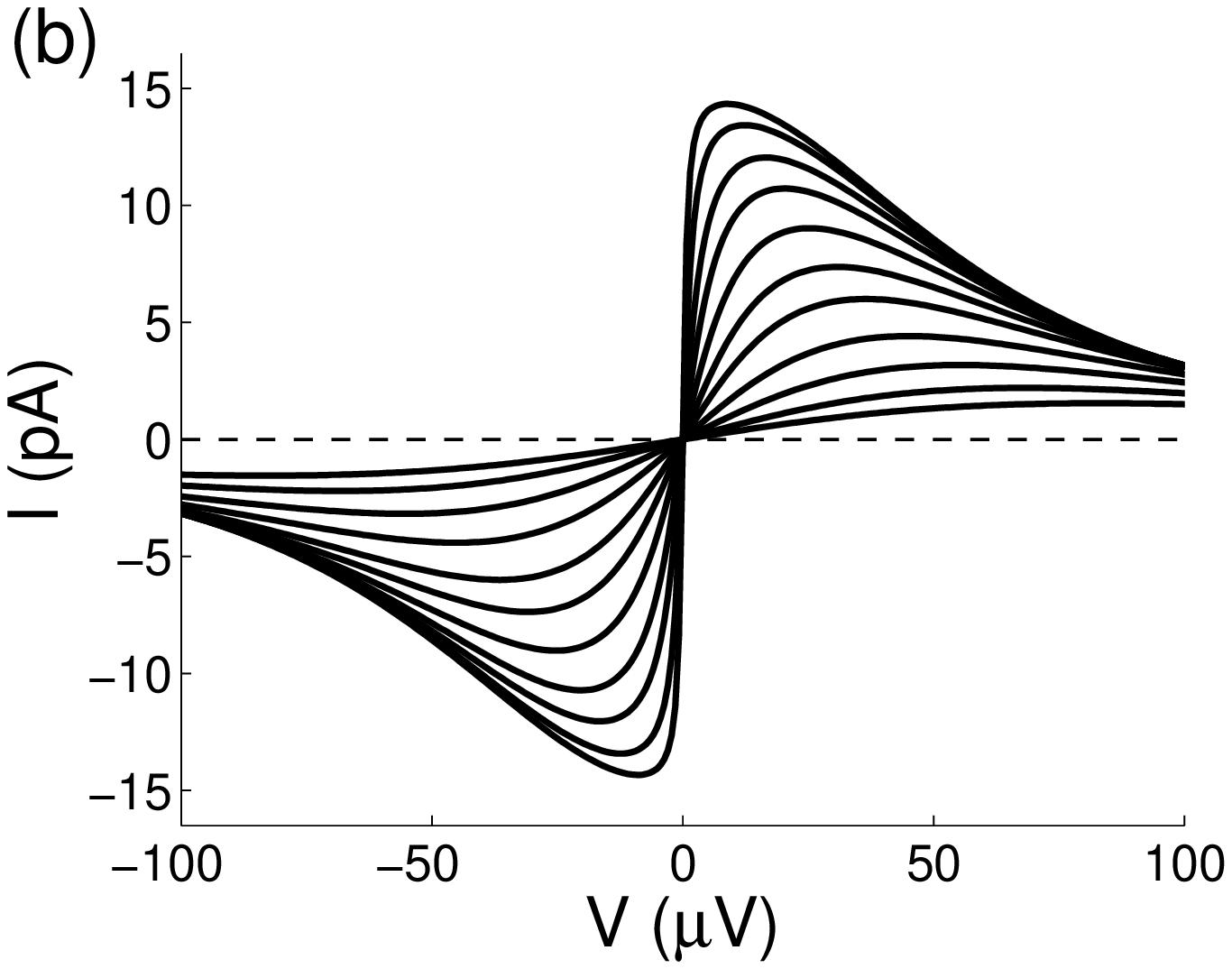}
\caption{(a) Experimentally measured I-V characteristics at $d/l_B=1.50$ and different temperatures $T=: 25, 50, 75, 100, 125, 150, 175, 200, 225, 250, 275$ mK.
(b) I-V characteristics obtained using Eq.~(\ref{altapp2}) for $V_0=73$ $\mu$V, $I_0=4.9$ pA and different values of $\alpha$, which are fitted at each temperature. The obtained temperature dependence of $\alpha$ is shown in Fig.~\ref{fig:alpha}. The experimental curves are reproduced from the experimental data provided by I. B. Spielman \textit{et al.} and reported in Ref.~\onlinecite{Spielman-diss}.}
\label{spielman53}
\end{figure}

\begin{figure}
\includegraphics[scale=0.5]{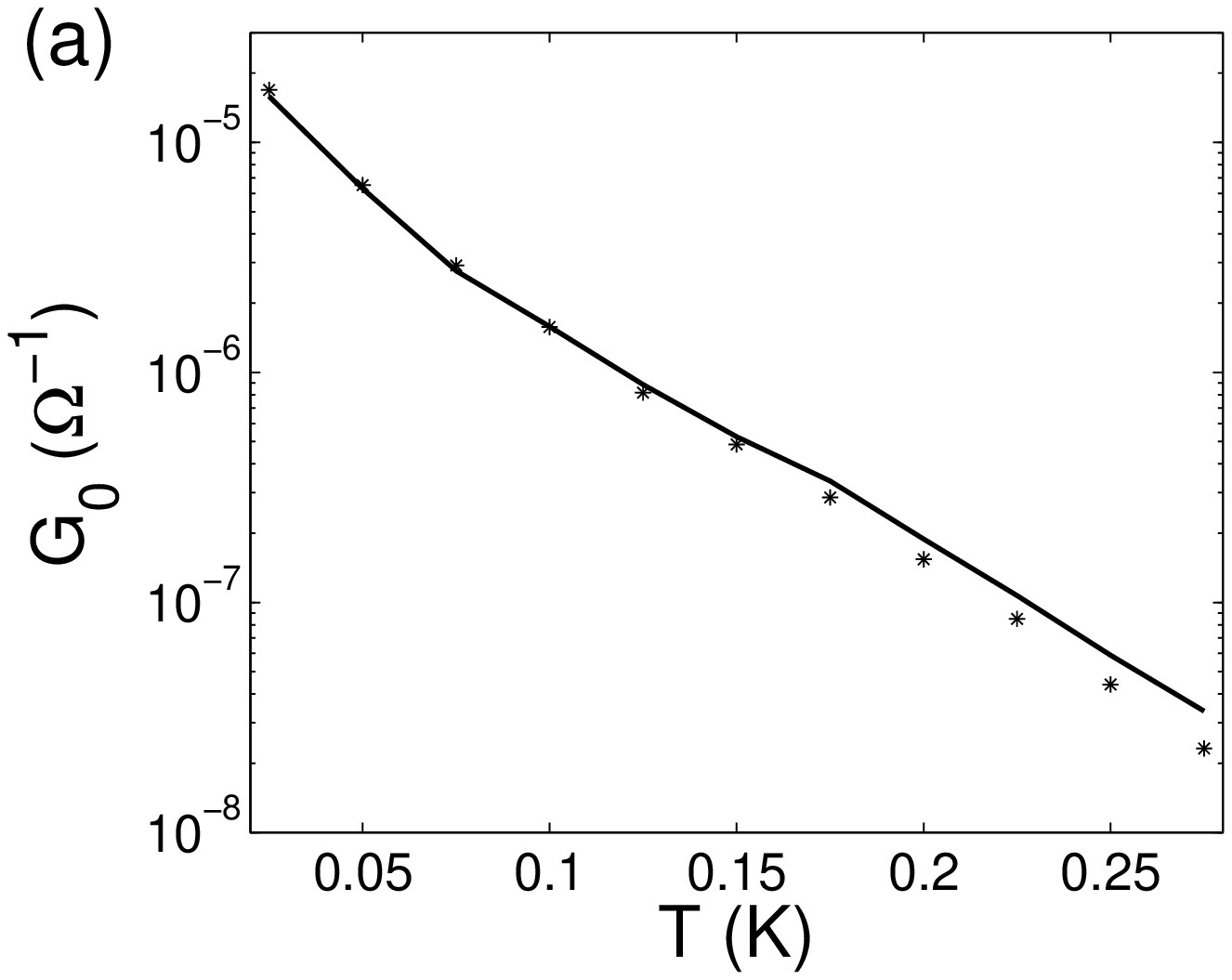} \includegraphics[scale=0.5]{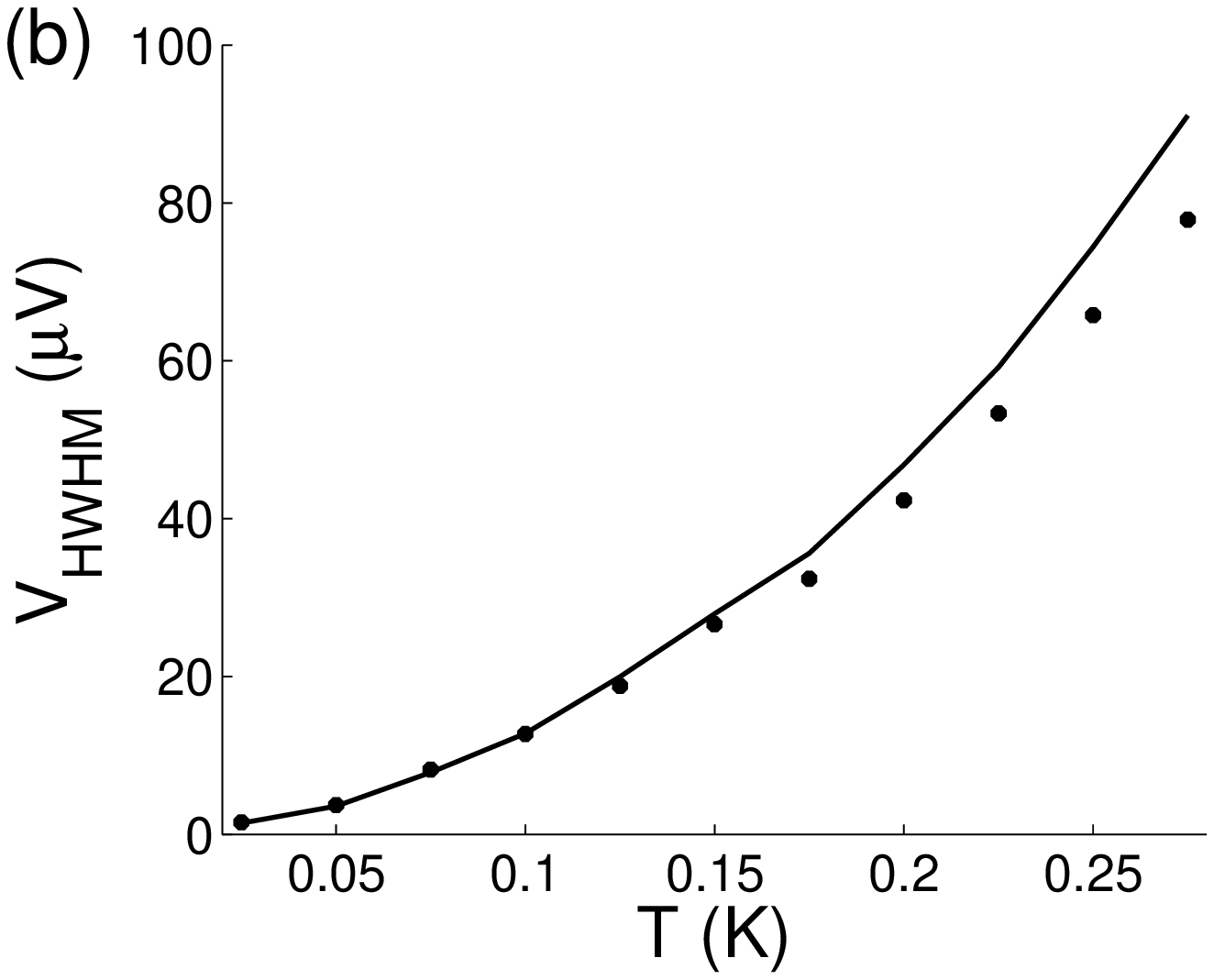}
\caption{Temperature dependencies of (a) the height and (b) the width of the conductance peak. Dots show the experimental measurements and the lines show the theoretical values obtained using $V_0=73$ $\mu$V, $I_0=4.9$ pA and the temperature dependence of $\alpha$ shown in Fig.~\ref{fig:alpha}. The experimental  data was provided by I. B. Spielman \textit{et al.} and has been reported in Ref.~\onlinecite{Spielman-diss}.}
\label{theoT2}
\end{figure}

\begin{figure}
\includegraphics[scale=0.58]{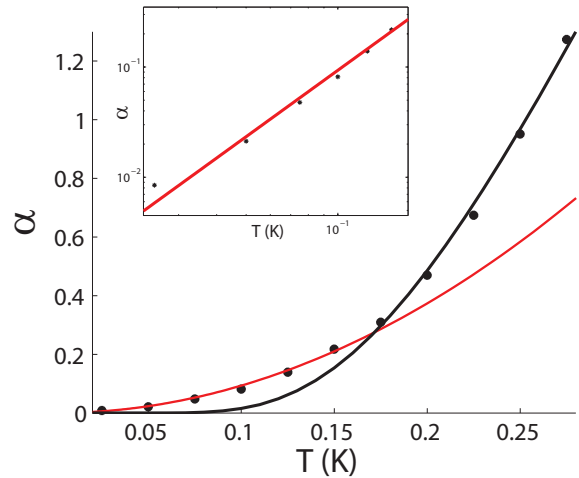}
\caption{(color online). Temperature dependence of $\alpha$. The dots show the temperature dependence of $\alpha$ obtained by fitting the theoretical height and width of the conductance peak to the experimental results. The black and red lines are guides to the eye. The black line is an exponential curve $\alpha=c \cdot \exp(-T_0/T)$ ($c \approx 15.3$ and $T_0 \approx 0.69$ K), where $c$ and $T_0$ are fitted using the high-temperature ($T\geq 150$ mK) experimental data. The red line shows $\alpha=(T/T_0)^2$ ($T_0 \approx 0.33$ K), where $T_0$ is fitted using the low-temperature ($T\leq 125$ mK) experimental data. Inset shows the temperature dependence of $\alpha$ at small temperatures plotted on a double logarithmic scale.}
\label{fig:alpha}
\end{figure}

The measured temperature dependence of the I-V characteristic is shown in Fig.~\ref{spielman53} (a).
Plotted on a single logarithmic scale, the temperature dependence of the peak conductance seems to
be consistent with $G_0 \propto \exp(- T/T_0)$ as pointed out in Ref.~\onlinecite{Spielman-diss} (see Fig.~\ref{theoT2}).
Such a  temperature dependence is not expected because it differs significantly from the temperature dependence of the tunneling current in standard Josephson junctions \cite{Steinbach, IvanchenkoZilberman}, and it seems naively that the small-bias tunneling conductance would stay finite even at  zero temperature \cite{Spielman-diss} in contrast to theoretical expectations \cite{Stern01, Balents01, Fogler-Wilczek01}.

We argue that the experimentally observed temperature-dependence of the I-V characteristic
can be described using the parametrization of our theory.  Based on the in-plane magnetic field dependencies of the small-signal tunneling conductance measured at different temperatures \cite{Spielman-diss}, we can estimate that $\xi$ depends only weakly on temperature. Moreover, we expect that $\rho_s$ and $\Gamma$ do not depend strongly on temperature at the experimentally relevant temperatures deep inside the coherent phase. This assumption could be tested by measuring the temperature dependence of $u$. Therefore, we assume that the main temperature dependence of the tunneling current must originate from the temperature dependence of $\tau_\varphi$. Since the tunneling current according to Eq.~(\ref{altapp2}) depends on $\tau_\varphi$ only through the parameter $\alpha$, we will use $\alpha$ as the fitting parameter.

Fig.~\ref{spielman53} (b) shows the theoretical I-V characteristics obtained using Eq.~(\ref{altapp2}). The values of $V_0$ and $I_0$ are assumed to be temperature independent 
and the parameter $\alpha$ is determined at each temperature as the average of the two values obtained by fitting the height and width of the conductance peak. The parameter $V_0=73$ $\mu$V is determined from  $\xi$ and $u$ estimated in previous section, and $I_0$ is fitted to optimize the agreement between the theoretically calculated and experimentally measured height and width of the conductance peak at low temperatures \cite{Fitdetail}.
The value of $I_0 =4.9$ pA obtained by fitting the theory to the experimental data is consistent with the estimate for $I_0 \sim 3.4-13.6$ pA (corresponding to $\Delta_{SAS}=5-10$ $\mu$K and $d/l_B=1.5$) based on Eq.~(\ref{I0exp}).
The theoretically obtained I-V characteristics are in quantitative agreement with the experimental ones. In particular, as demonstrated in Fig.~\ref{theoT2}, we can quantitatively obtain the temperature dependencies of both the height and the width of the conductance peak by fitting a single parameter $\alpha$ at each temperature. Therefore, the approach outlined above is justified and we can concentrate on the temperature dependence of $\alpha$, which is shown in Fig.~\ref{fig:alpha}. We can separate two regimes of temperatures, where the temperature dependence of $\alpha$ appears to be  different from each other. As demonstrated in Fig.~\ref{fig:alpha}, at small temperatures it is conceivable that $\alpha$  obeys a power law,
whereas at high-temperatures $\alpha$ shows thermally activated behavior $\alpha \propto e^{-T_0/T}$.  These results seem to be in agreement with the expected temperature dependencies of the different physical processes possibly contributing to vortex field fluctuations \cite{Fertig03, Fertig05, Huse05, Roostaei08, Eastham09}. The large temperature behavior could be attributed to thermally activated hopping of merons whereas the power-law dependence could result from localized low-energy excitations \cite{Fertig03}. Importantly, our analysis shows that as $T \to 0$ our parameter $\alpha \to 0$ indicating that $\tau_\varphi \to \infty$. (At lowest temperatures $\hbar/\tau_\varphi < k_B T$ and it  is still decreasing reasonably rapidly with decreasing temperature.)

According to Eq.~(\ref{zerobiasG}),
 the conductance diverges in the limit $\tau_\varphi \to \infty$. Therefore, within  our approach a diverging conductance at small temperatures is still expected by extrapolating the temperature dependence of $\alpha$, which is consistent with the experimentally measured temperature dependencies of the height and width of the conductance peak. According to this  analysis, the apparent temperature dependence $G_0 \propto \exp(- T/T_0)$ of the of the small-signal conductance arises because two different physical processes are contributing to the vortex field fluctuations at low and high temperatures, respectively.

We also point out that $\tau_\varphi$ can, in principle, be determined experimentally using a time-dependent interlayer voltage. Similarly to the case of Josephson junctions in the presence of strong fluctuations \cite{Shapiro-like-steps}, we expect that an ac field induces structures in the I-V characteristic at the resonances $eV \approx n \hbar \omega$ ($n=1,2...$). These resonant structures resemble the Shapiro steps, but they are distinct from each other only if the frequency of the ac field satisfies $\omega > 1/\tau_\varphi$, allowing independent determination of $\tau_\varphi$ at different temperatures. Alternatively $\tau_\varphi$ could be determined by measuring the frequency dependence of the dynamical conductivity. At the smallest temperatures the required frequencies of the ac field are in the GHz frequency range.

\section{Qualitative changes caused by larger tunneling amplitude \label{largetun}}

We next study the effects caused by a larger tunneling amplitude. In a series of recent tunneling experiments \cite{TiemannNJP08, TiemannPRB09, Tiemann-diss, Yoon} samples with otherwise similar parameters but significantly larger tunneling amplitudes were used. The tunneling amplitude can again be determined either by using the tunneling I-V characteristics in the absence of magnetic field or by solving the Schr\"{o}dinger and Poisson equations self-consistently. The obtained values of $\Delta_{SAS}$ vary between $100-150$ $\mu$K \cite{TiemannPRB09, Tiemanndata}. This good agreement between the values obtained using two different methods suggests that the estimates for $\Delta_{SAS}$ are reliable.
In comparison with the experiments discussed in the previous section, the tunneling amplitude of the samples in these experiments is therefore approximately an order of magnitude larger. Based on the scaling law $I_0 \propto \Delta_{\textrm{SAS}}^2$, Eq.~(\ref{I0exp}), we expect that  the tunneling currents are approximately two orders of magnitude larger $I_0 \sim 0.5-1$ nA \cite{comment}. This kind of scaling is in agreement with the experimental observations \cite{TiemannNJP08, TiemannPRB09, Tiemann-diss, Yoon}. Moreover, it was observed in the experiments \cite{TiemannPRB09} that the tunneling current is still proportional to the area of the sample. These observations indicate that essentially the same fluctuations-dominated physics determines the tunneling I-V characteristics also in these experiments. On the other hand, there are several important qualitative differences, which are discussed below. These new effects arise because the tunneling resistance is no longer large compared to the other resistances in the system.

\subsection{ Tunneling geometry}

Because $I_0$ is now two orders of magnitude larger than in the earlier experiments, we estimate that the tunnel resistance at low temperatures is  comparable to the contact resistances and intralayer resistances $R_{xx}$ and $R_{xy}$. Although the interlayer voltage $V(\vec{r})$ is no longer constant, we can still expect that it varies reasonably slowly in space. This means that we can describe the transport in this system by assuming that the local tunnel current density is $I(V(\vec{r}))/L^2$, where $V(\vec{r})$ is the local interlayer voltage and $I(V)$ is the current-voltage characteristic given by Eq.~(\ref{altapp2}). The interlayer voltage $V(\vec{r})$ can now be calculated by assuming that the intralayer transport and counterflow currents are described by resistivity tensors obtained from the experiments \cite{quantized-Hall-Drag1, quantized-Hall-Drag2, Kellog04, Tutuc04, Tutuc05, Wiersma04-06}. An effective transport theory under these assumptions can be formulated using the current continuity equation and appropriate boundary conditions, which are determined by the experimental geometry.

Here we do not attempt to explain quantitatively all the details of the experiments but instead we just want to qualitatively explain the main features observed in the I-V characteristics. The main effect of the inhomogeneities in $V(\vec{r})$ is that all the sharp features in the I-V characteristics become rounded and dependent on the locations of the voltage probes. In particular, it becomes difficult to determine the tunneling conductance at small voltages, because the tunneling resistance is small compared to the other resistances in the system and the observed
values of the interlayer
voltage can depend on the locations of voltage probes \cite{Tiemann-diss}.
Nevertheless, the temperature dependence of the peak tunneling current \cite{TiemannPRB09, Tiemann-diss}  is similar to that observed  in the earlier experiments (Section \ref{sec:tempdep}) and therefore we expect that the perturbative approach is at least approximately valid even at the lowest experimental temperatures. By keeping in mind the type of small modifications discussed above, we can mainly ignore the inhomogeneities in the analysis of the remaining effects caused by larger tunneling amplitude.

\begin{figure}
\includegraphics[scale=0.4]{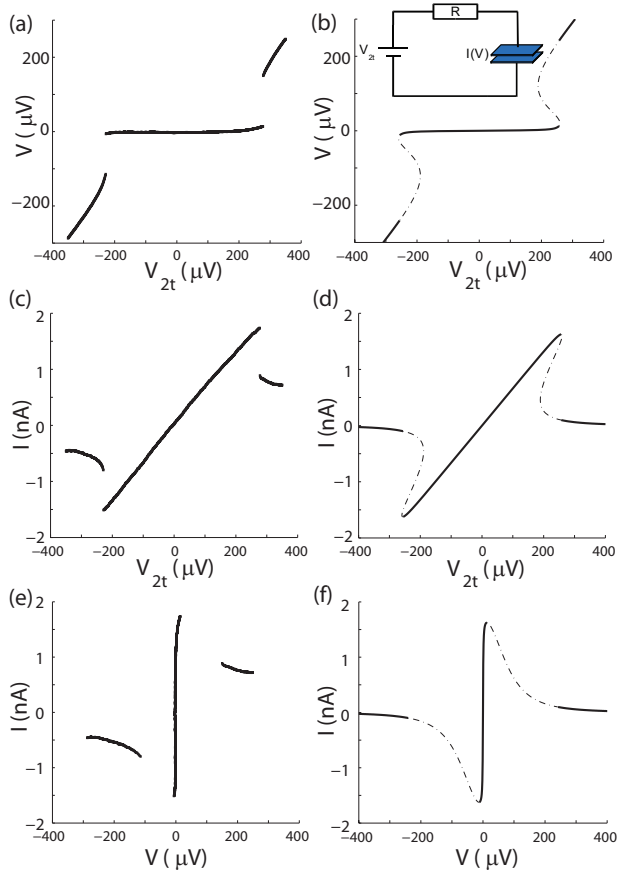}
\caption{(color online). (a),(c) Experimentally measured dependencies of the interlayer voltage $V$ and the tunneling current $I$ on the applied voltage $V_{2t}$. (e) Experimentally measured tunneling current $I$ as a function of interlayer voltage $V$. The experimental measurements were made at $d/l_B=1.42$ and $T<20$ mK. (b), (d), (f) Corresponding theoretically calculated tunneling characteristics obtained using Eqs.~(\ref{circeq}) and (\ref{altapp2}). The theoretical parameters used in the calculation are $\alpha=0.01$, $V_0=100$ $\mu$V and $I_0=0.56$ nA. The inset in figure (b) shows the effective circuit. Resistance $R$ is caused by the circuit and contact resistances. There exists a range of voltages $V_{2t}$, where Eq.~(\ref{circeq}) has several solutions. The additional solutions are indicated by the thin dashed lines in figures (b), (d) and (f). The experimental curves are reproduced from the experimental data provided by L. Tiemann \textit{et al.} and reported in Refs.~\onlinecite{TiemannPRB09} and \onlinecite{Tiemann-diss}.}
\label{stuttgarttemp}
\end{figure}

Figs.~\ref{stuttgarttemp}(a), (c) and (e) show the results of experimental measurements at $d/l_B=1.42$ and $T<20$ mK, reported in Ref.~\onlinecite{TiemannPRB09}. The I-V characteristic shown in Fig.~\ref{stuttgarttemp}(e) is qualitatively very similar to the I-V characteristics obtained for samples with smaller tunneling amplitude (see Fig.~\ref{spielman53}). The main difference to the earlier experiments is that there now exists jumps at particular values of the voltage. If we assume that the inhomogeneities are not important, the experimental geometry can be described with an effective circuit shown in the inset in Fig.~\ref{stuttgarttemp} (b). This gives us a circuit equation
\begin{equation}
V_{2t}-RI=V(I), \label{circeq}
\end{equation}
where the $V(I)$ characteristic of the bilayer tunnel junction is obtained from Eq.~(\ref{altapp2}) by solving the voltage $V$ as a function of current $I$. The solutions of Eq.~(\ref{circeq}) are obtained by finding the intersections of $V(I)$ characteristic and the lines $V_{2t}-RI$. We assume that $\alpha=0.01$ and $V_0=100$ $\mu$V in agreement with the values found in Section \ref{Caltech}. Because the effective resistance $R$ is significantly larger than the tunneling resistance at small voltages, the resistance $R$ can be determined from Fig.~\ref{stuttgarttemp}(c) by calculating the slope of the $I(V_{2t})$ characteristic at small applied voltages $V_{2t}$. We obtain $R \approx 150$ k$\Omega$. Moreover, by fitting the theoretically calculated maximum of the tunneling current to the experimentally observed maximum current, we obtain $I_0=0.56$ nA. This value is in good agreement with an estimate $I_0 \sim 0.65$ nA obtained using Eq.~(\ref{I0exp}) for $d=28.6$ nm, $d/l_B=1.4$, $L_x L_y=880\cdot 80$ $\mu$m$^2$, $\Delta_{SAS} = 100$ $\mu$K and $\xi=92$ nm (corresponding to $V_0=100$ $\mu$K). Here we have assumed that $u$, $\rho_s$ and $\Gamma$ are the same as in the Caltech experiments.
The results of the theoretical calculations for these values of $\alpha$, $R$ and $I_0$ are shown in Figs.~\ref{stuttgarttemp} (b), (d) and (f). They are in good agreement with the experimental measurements. Moreover, we find that for particular values of $V_{2t}$, there exists several solutions of Eq.~(\ref{circeq}) indicating bistability. We expect that the jumps in the measured I-V characteristic [see Fig.~\ref{stuttgarttemp} (e)] originate from a hysteretic switching caused by this bistability.

\subsection{Counterflow geometry}

\begin{figure}
\hspace{-0.7cm}\includegraphics[scale=0.63]{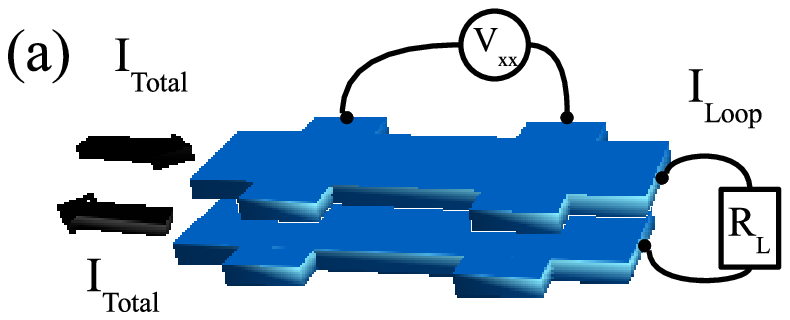}
\includegraphics[scale=0.26]{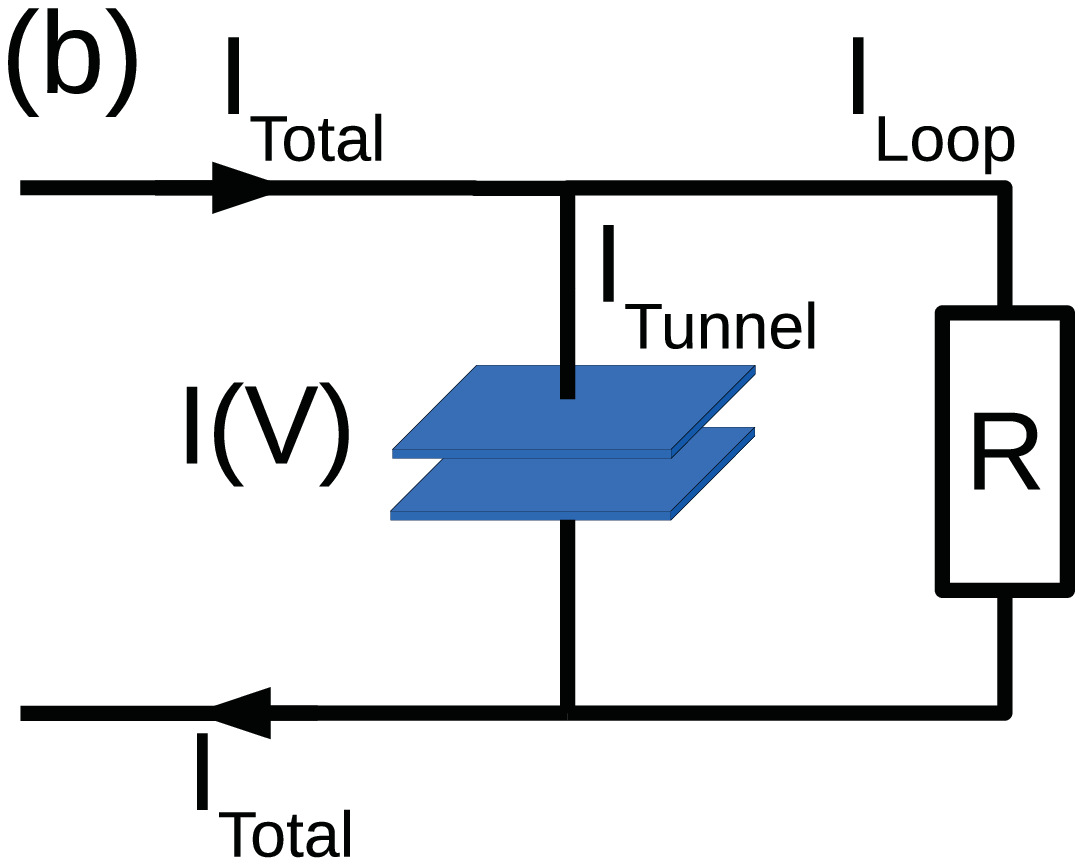}
\includegraphics[scale=0.5]{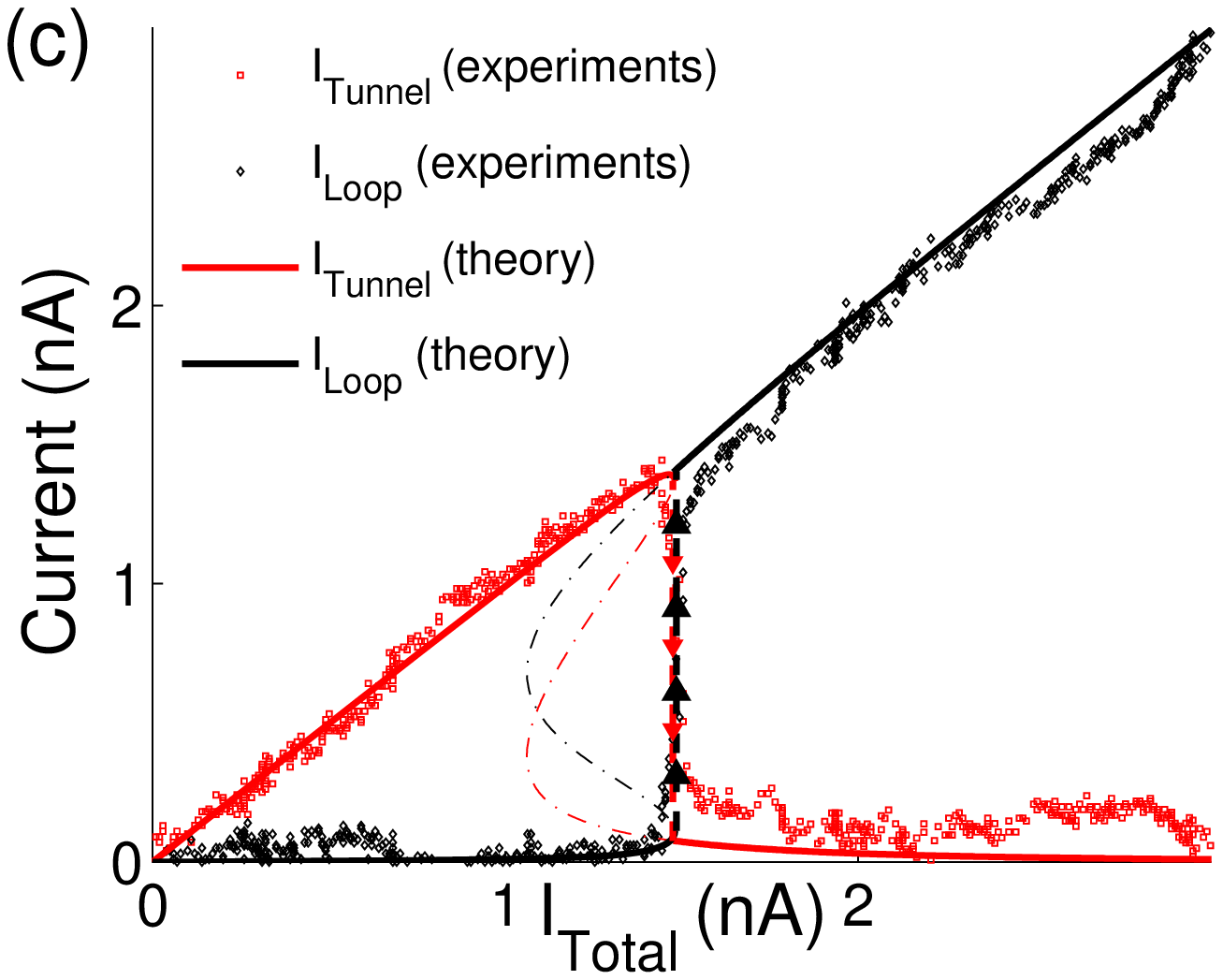} \includegraphics[scale=0.5]{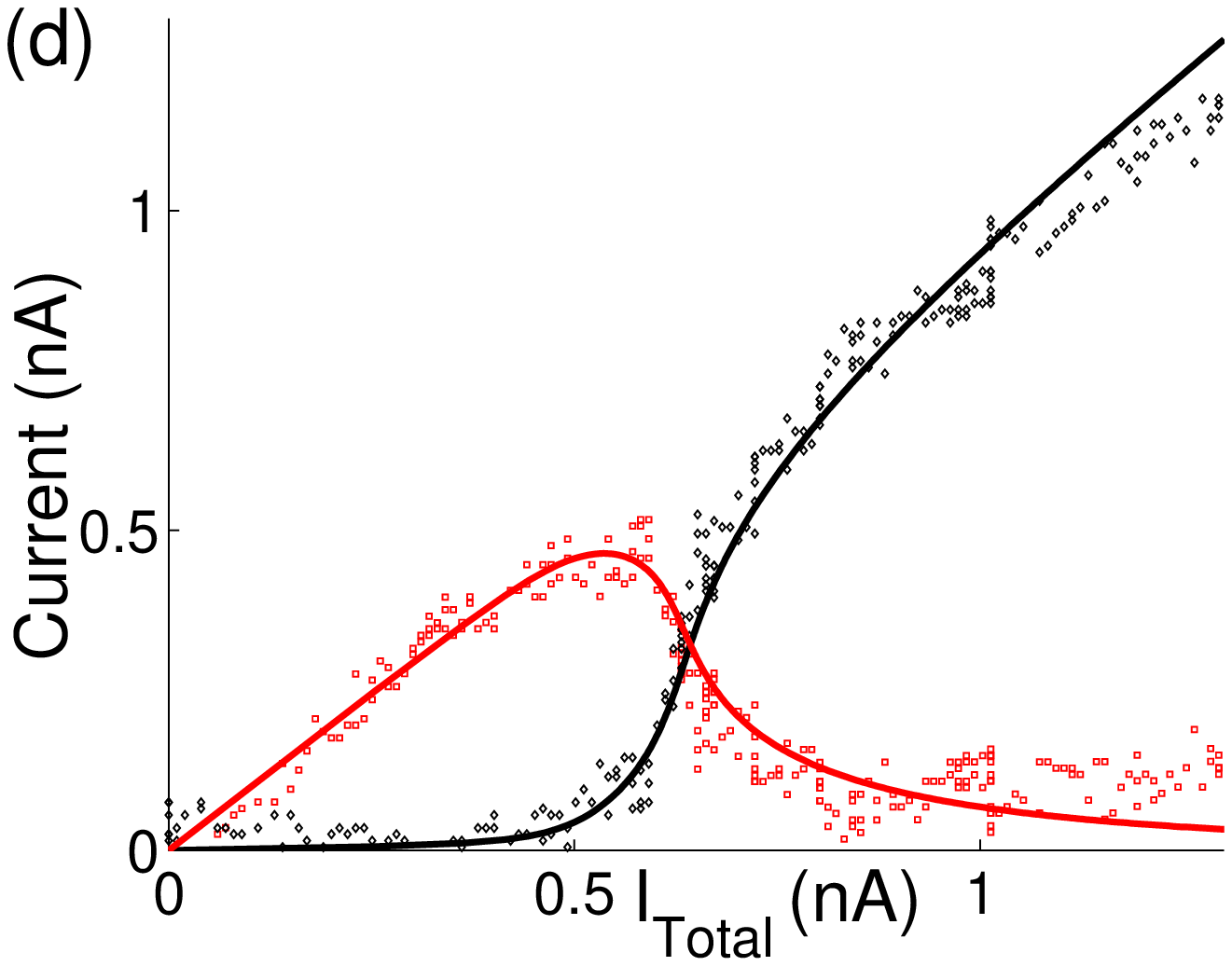}\vspace{0.4 cm}
\includegraphics[scale=0.27]{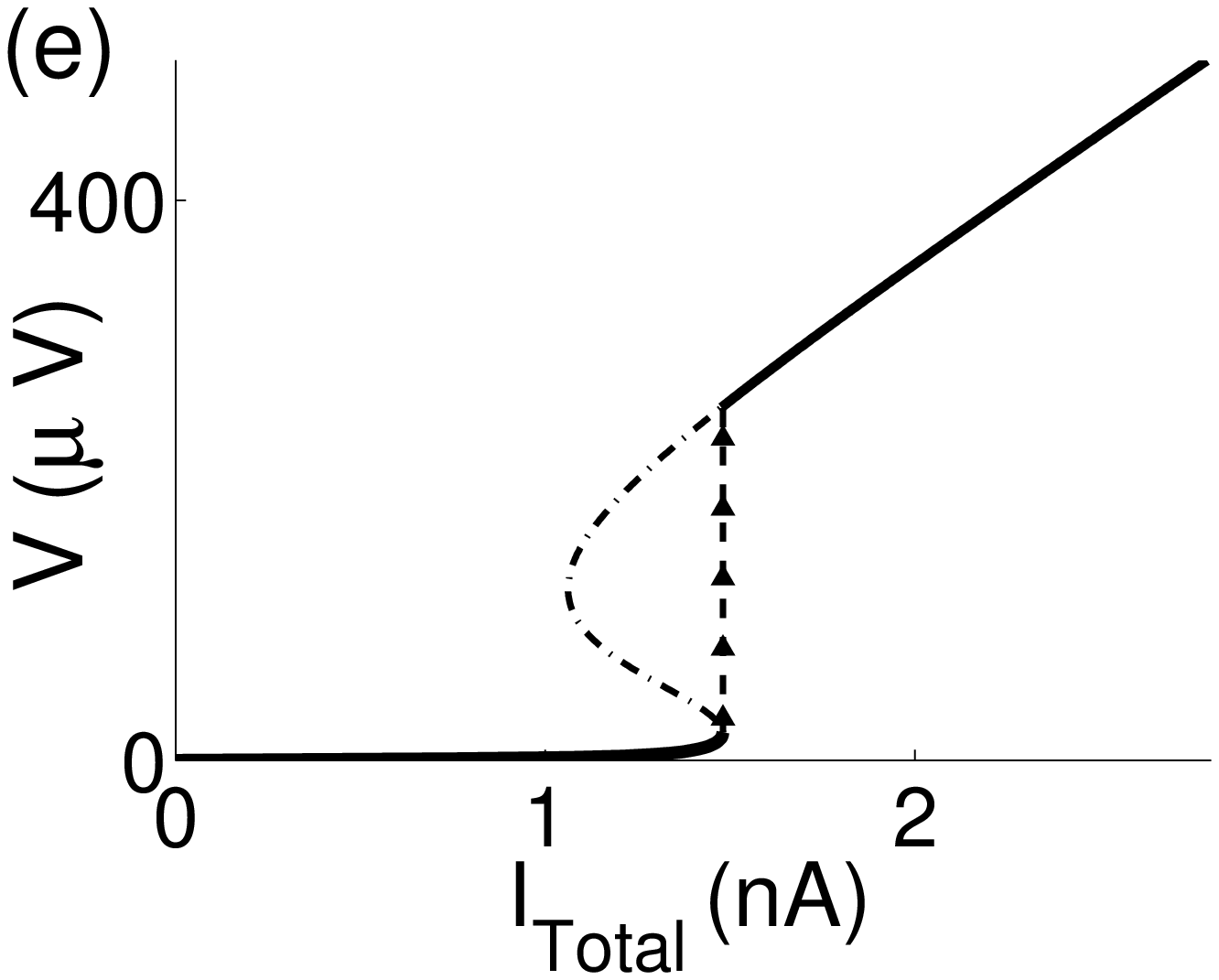}  \includegraphics[scale=0.27]{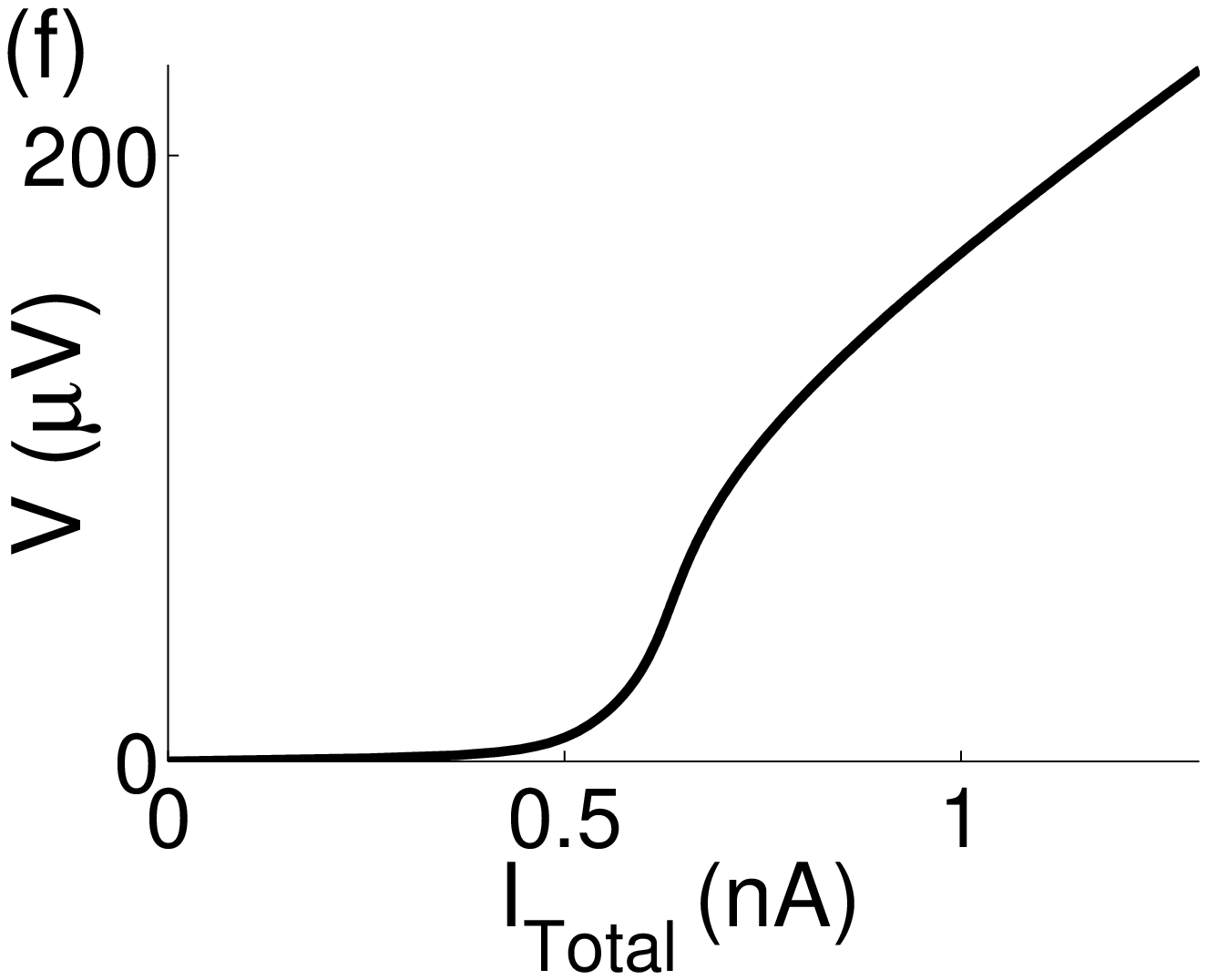}
\caption{(color online). (a) Counterflow geometry. (b) Equivalent circuit assuming homogeneous interlayer voltage. Here $R$ is an effective resistance, which takes into account the loop resistance and contact resistances. (c), (d) Tunneling and loop currents as a function of total current. The experimental results shown in figures (c) and (d) are obtained for $d/l_B=1.37$ and $d/l_B=1.68$, respectively.  In theoretical calculations we assume that $R$, $\alpha$ and $V_0$ are independent of $d/l_B$. The parameter $I_0$ is fitted to be $I_0=0.48$ nA and  $I_0=0.16$ nA for the different values of $d/l_B=1.37$ and $d/l_B=1.68$, respectively. (e), (f) Interlayer voltages corresponding to the parameters used in figures (c) and (d), respectively. For $I_0=0.48$ nA there exist an interval of $I_{\textrm{Total}}$, where the Eq.~(\ref{circ2}) has several solutions. The additional solutions are shown by the thin dashed lines in figures (c) and (e). The experimental curves are reproduced from the experimental data provided by Y. Yoon \textit{et al.} and reported in Ref.~\onlinecite{Yoon}.}
\label{counterflowcurtheo}
\end{figure}

Recently, tunneling experiments were performed also in the counterflow geometry [Fig.~\ref{counterflowcurtheo} (a)] using samples with similar $\Delta_{SAS} \sim 100-150$ $\mu$K.\cite{Yoon}  In this case, dc input current is generated by applying a voltage to a large resistor in series with the system, and the current which passes through the top layer is redirected via a loop resistor into the bottom layer. If we assume that  inhomogeneities are not playing an important role, we can construct an equivalent circuit shown in Fig.~\ref{counterflowcurtheo}(b). The circuit equation is now
\begin{equation}
I_{\textrm{Total}}=I_{\textrm{Tunnel}}+I_{\textrm{Loop}}=I(V)+V/R, \label{circ2}
\end{equation}
where $I(V)$ is the tunneling I-V characteristic of the bilayer and $R$ is an effective resistance, which takes into account the loop resistance $R_L$ and the contact resistances.

Two intriguing observations were made in the experiment reported in Ref.~\onlinecite{Yoon}.
It was found that the tunneling current was large as long as the injected drive current was smaller than a certain critical current $I_c$ at which point
a transition into a new regime with negligible tunneling current was observed.  For small $d/l_B$ the observed transition is extremely sharp and becomes smoother for larger $d/l_B$ as shown in Figs.~\ref{counterflowcurtheo} (c) and (d), respectively.  Secondly, it was found that although the transition between the two tunneling regimes had only a minor impact on the state as observed in the magnetotransport, the longitudinal resistance increased rapidly for currents exceeding the critical current $I_c$. Our theory provides a simple and natural explanation for both experimental observations.

In order to explain these observations we assume that $\alpha=0.01$  and $V_0=100$ $\mu$V similar to the values used in previous sections. We determine
$R=180$ k$\Omega$
and  $I_0 = 0.16$ nA ($d/l_B=1.68$) and $I_0=0.48$ nA ($d/l_B=1.37$) such that we visually obtain
good agreement between theoretical and experimental dependencies of loop and tunneling current on the
total current. The values of $I_0$ are in agreement with the measured tunneling currents at different $d/l_B$ in the tunneling geometry \cite{TiemannPRB09, Tiemann-diss}. The effective resistance $R$ is significantly larger than the loop resistance $R_L=10$ k$\Omega$ used in the experiment \cite{Yoon}, and therefore we expect that small variations in $R_L$ would not affect the tunneling and loop currents significantly.  Our calculations show that depending on the value of $I_0$ there exist two qualitatively different behaviors similarly to the ones observed in experiments. By inspecting the tunneling I-V characteristics (Figs.~\ref{spielman53} and \ref{stuttgarttemp}), we see that after the peak tunneling current has been reached the tunneling current decreases quickly with increasing voltage. If $I_0/V_0$ is large enough this decrease is more rapid than the increase of the loop current through the resistor.
In this case, there exist several solutions of Eq.~(\ref{circ2}) around the critical current as demonstrated with the help of dashed lines in Figs.~\ref{counterflowcurtheo} (c), (e). We expect that this results in bistability, which can be seen as abrupt changes in the tunneling and loop currents in the experiments at small $d/l_B$. At large $d/l_B$ the ratio $I_0/V_0$ is expected to be smaller and therefore the change in the tunneling and loop currents becomes  smoother near the critical current.
Our theoretical picture explains naturally why the transition has only a minor impact on the $\nu_T=1$ state as observed in  magnetotransport \cite{Yoon}. The reason is simply that the transition occurs due to the tunneling I-V characteristic, which is an intrinsic property of the $\nu_T=1$ state, and therefore in the first approximation it should have no effect at all on the $\nu_T=1$ state. Moreover, according to our theory it is natural that the critical current observed in this counterflow geometry is almost exactly the same as the peak tunneling current observed in the tunneling geometry.

As already pointed out, the experiments also indicate that the dissipation increases rapidly at the currents exceeding the critical current. It has been suggested \cite{Yoon} that the increase of dissipation could be caused either by merons \cite{Fertig03, Fertig05, Huse05, Roostaei08, Eastham09} or so-called Josephson vortices \cite{Fil09}. However, the Josephson vortices could only exist if the Josephson length $\lambda_J=\sqrt{4 \pi l_B^2 \rho_s/\Delta_{SAS}}$ was smaller or at least comparable to the coherence length $\xi$. This is not the case in the present experiments where we estimate $\lambda_J \sim 4$ $\mu$m. Moreover, the existence of Josephson vortices would indicate significantly larger critical currents \cite{Fil09} than observed in the experiments. Therefore we expect that merons are responsible for the increased dissipation.

In our theoretical picture, the increase of dissipation should be due to a change in the dynamics of merons, which can be due
to either an abrupt change in interlayer voltage or due to large counterflow current.
 If the effect was caused by the counterflow current it seems difficult to understand why the critical current is almost exactly equal to the maximum current observed in the tunneling geometry. Moreover, by assuming that the counterflow current induces a Magnus force on merons \cite{Huse05} and that merons are confined into potential wells corresponding to the measured thermal activation gap \cite{Wiersma04-06}, we estimate that the critical counterflow current necessary to depin the merons is significantly larger than typical experimental currents.
 On the other hand, we find that for small $d/l_B$ the interlayer voltage significantly increases at the critical value of total current. Moreover, for both values of  $d/l_B$, the interlayer voltage increases rapidly with increasing total current for currents exceeding the critical current [Figs.~\ref{counterflowcurtheo} (e), (f)], resembling the experimentally observed dependence of the dissipation on the total current \cite{Yoon}. Therefore, we believe that the increasing dissipation is tightly connected with the increasing interlayer voltage. By using our estimate for the exchange enhanced capacitance $\Gamma$, we find that an interlayer voltage of $100$ $\mu$V can result in a density imbalance of several percents. It has been experimentally observed \cite{Wiersma04-06} and theoretically explained \cite{Roostaei08} that such a density imbalance can significantly affect the longitudinal resistance. Furthermore, it is possible to experimentally test this hypothesis, because the density imbalance would result in different longitudinal resistances in the upper and lower layers \cite{Wiersma04-06, Roostaei08}.

\section{Relation to previous theoretical approaches \label{others}}

The theoretical explanation of charge  transport in bilayer quantum Hall systems has been a long-standing theoretical problem. Perhaps most surprisingly, instead of a true Josephson effect, a Josephson-like conductance peak has been observed experimentally.  Understanding the finite height and width of this small bias conductance peak and the calculation of critical tunneling currents has attracted a
significant amount of  interest. Theoretical approaches differ in whether they take a clean or a disordered system as a starting point of their analysis.

In the clean limit, the critical current is determined as the maximum current for which the pseudospin field can be static, and the dynamics of charge transfer between leads and bulk by quasi-particles plays an important role.
Conductance peak
heights and maximum tunnel currents predicted  in the clean limit   are orders of magnitude larger than the values observed in experiments,\cite{Su10} and also differ from experiments with respect to  the parametric dependencies of conductance
peak and maximum tunnel current on
sample area,  bare tunneling amplitude,   and  in-plane magnetic field. Most notably, the critical tunnel current
depends linearly or sublinearly on the  tunneling amplitude $\Delta_{SAS}$, and is independent of system size for
systems larger than the Josephson length $\lambda_J$.
In contrast, experiments seem to be consistent with a critical tunnel current proportional to both the area of the
sample and to the square of the tunneling amplitude.

The influence of quenched disorder can be described by the introduction of a vortex field, which causes an exponential
decay of pseudospin correlations with a characteristic spatial correlation length $\xi$. Under the assumption of a
static vortex field,  local pseudospin correlations  are  assumed to have an infinite range in time.
One-dimensional numerical simulations  indicate that the static vortex field results in an intriguing self-organized Bean critical state,  where the current injected at the boundaries can affect the condensate dynamics throughout the whole sample.\cite{Eastham10} By generalizing this finding to two-dimensional system, the authors in Refs.~\onlinecite{Eastham10} and \onlinecite{LeeEC10} find that if the sample size is sufficiently large, the critical tunneling current is proportional to the area of the sample. This statement can be justified with the help of an energy optimization argument in the presence of random static vortex field \cite{Eastham10, LeeEC10}, and it turns out that the sample size should be compared to a length scale $\lambda_J^2/\xi$. In the experiments reported in Ref.~\onlinecite{TiemannPRB09} the conditions $L_x,L_y > \lambda_J^2/\xi$ are approximately satisfied and the critical tunneling current proportional to the area of the sample has been observed. On the other hand, the approach based on the assumption of a static vortex field does not
easily explain  why a similar scaling with the area has been observed \cite{Finck08} also in samples, where the sample size is not large enough in comparison to the Josephson length.
Other challenges for this type of approach are the explanations of the experimentally observed temperature and parallel magnetic field dependencies of the tunneling current, and the quantitative description of the height and width of the conductance peak. Nevertheless, one would expect that the  vortex field should
be static in the limit of zero temperature, and that the predictions of Refs.~\onlinecite{Eastham10} and \onlinecite{LeeEC10} should
describe the limit of extremely low temperatures, which probably has not yet  been reached experimentally.

Effects of a dynamical vortex field were analyzed in Refs.~\onlinecite{Stern01} and \onlinecite{Balents01}. In the present manuscript,
we follow the approach outlined in Ref.~\onlinecite{Stern01},  introduce a phenomenological vortex field, and  calculate the dependence of experimental observables on the correlation length $\xi$ and the correlation time $\tau_\varphi$ of the vortex field. We argue that all the important temperature dependence of experimental observables originates in the temperature
dependence of the dimensionless decoherence rate $\alpha(T) = \xi /(u \tau_\varphi(T))$, and we determine  $\alpha(T)$
by fitting experimental curves. The dynamics of merons was explored in Ref.~\onlinecite{Fertig03} by means of mapping
 the quantum Hall bilayer to a classical two-dimensional XY model with a symmetry-breaking field and with disorder. It appears that
 disorder nucleates strings of overturned spins, which connect vortices and antivortices at their ends. At low temperatures, this state is characterized by anomalously large fluctuations of the vortex field, which are reminiscent of glassy dynamics.
We believe that the power law like, weak temperature dependence of $\alpha$ which we extract from experiment may be
consistent with the results of Ref.~\onlinecite{Fertig03}.

\section{Discussion and Conclusions \label{conclusions}}

In this paper we have studied how the Josephson-like tunneling depends on the area and tunneling amplitude of the samples, applied parallel magnetic field and temperature. We have compared  different theoretical approaches describing the Josephson-like tunneling and found that all the tunneling experiments are in agreement with a theory which treats fluctuations due to meron excitations phenomenologically and takes tunneling into account perturbatively. Previously,  Josephson-like tunneling in samples with smaller \cite{Spielman00, Spielman01, Eisenstein03, Spielman04, Spielman-diss, Spielman05, Finck08, Champagne08, Champagne08PRB}  and larger \cite{TiemannNJP08, TiemannPRB09, Tiemann-diss, Yoon} tunneling amplitudes have often been discussed in different terms, because two orders of magnitude larger tunneling currents are found in latter experiments. Nevertheless, our analysis shows that all these experiments can be discussed within a common framework, and the observed magnitudes of the tunneling currents are in good agreement with the theory. We have also shown that the somewhat surprising temperature and in-plane magnetic field dependencies of tunneling current can be explained quantitatively by describing the effects caused by the merons with the help of a phenomenological vortex field.
In this paper, we have concentrated on the description of the tunneling deep inside the coherent phase allowing us to make several simplifying approximations in Section \ref{theory}. However, we note that in principle the equations (\ref{main1}), (\ref{main2}), (\ref{main3}) and (\ref{disordercor}) could also be used to describe the Josephson-like tunneling closer to the phase-boundary separating the coherent and incoherent phases. This would allow the determination of the theoretical parameters of the model as a function of different experimentally controllable quantities such as $d/l_B$, density imbalance and temperature.

The surprisingly good agreement found between the theory and experiments naturally introduces several new theoretical questions. The temperature dependence of the tunneling currents was explained by fitting the parameter $\tau_\varphi$, which describes the dynamics of the vortex field. However, the details related to the microscopic origin of the vortex field dynamics are not well-understood. The exponential temperature dependence at large temperatures most likely results from the thermally activated hopping of the merons, but the mechanism giving rise to the power law dependence at low temperatures requires additional theoretical analysis.
Other interesting questions are related to the validity of the main approximation in our approach, which is the treatment of the tunneling as a perturbation. In general, the validity of this approximation depends on the nature of vortex-field fluctuations and dissipation mechanisms. We do not try to specify the dissipation processes here. Instead we expect on general grounds that the perturbative approach is controlled by a condition
\begin{equation}
\frac{L_i^2}{L^2} \frac{I(V)}{V} \ll \frac{e^2}{h}, \label{condit}
\end{equation}
where $L_i^2$ defines the size of a domain which is large enough such that its dynamics can be considered to be independent of the other domains. The idea behind this inequality is that the left side defines the conductance of an independent domain and we interpret it as a transmission probability multiplying $e^2/h$. If the condition (\ref{condit}) is satisfied, the transmission probability is much smaller than one and the higher order tunneling processes are suppressed by the powers of the transmission probability. There are two relevant length scales, $\xi$ and $u \tau_\varphi$, in the problem, which could determine the size of the independent domain $L_i$. Because at low temperatures $u\tau_\varphi \gg \xi$, the assumption $L_i^2 \approx (u \tau_\varphi)^2$ certainly yields the sufficient  condition
\begin{equation}
\frac{\xi^4}{\lambda_J^4} \frac{\rho_s}{\hbar/\tau_\varphi} \frac{\pi e^{-D_0/2}}{\alpha^2} \ll 1 \label{speccond}
\end{equation}
for the applicability of perturbation theory.
In Appendix \ref{calculation2}, we obtain another condition (\ref{parameter}) for the validity of the perturbative approach in terms of the dissipation strength $\gamma$. For dimensional reasons, we expect that $\gamma u^2 \propto 1/\tau_\varphi^2$ so that the only possible difference between conditions (\ref{speccond}) and (\ref{parameter}) is the reasonably large prefactor $\rho_s/(\hbar/\tau_\varphi)$ appearing in condition (\ref{speccond}). This prefactor essentially originates from our assumption for the area $L_i^2$ of the independent domains. The result of the classical calculation in Appendix  \ref{calculation2} can be interpreted as suggesting that the area for independent domains should be chosen as $L_i^2 \approx u \tau_\varphi \xi$, giving a less
restrictive criterion for the applicability of perturbation theory.

Because the assumption used to obtain Eq.~(\ref{speccond})  likely overestimates $L_i$, we consider the condition (\ref{speccond}) as a sufficient condition for the validity of the perturbative treatment of the tunneling, and we are confident that if it is satisfied, the perturbative approach is well justified. We find that the condition (\ref{speccond}) is always satisfied in the experiments with smaller tunneling amplitude. In the case of larger tunneling amplitude, assuming that the temperature dependence of $\alpha$ does not change, this condition could break down (depending on the exact values of different theoretical parameters) at the smallest experimental temperatures. The experimental observation \cite{TiemannNJP08, TiemannPRB09, Tiemann-diss, Yoon} that there is no clear transition to a different operation regime as a function of temperature could indicate that the necessary condition for the perturbative treatment of tunneling is less restrictive than the condition (\ref{speccond}) or that the larger tunneling currents resulted in an increase of the dissipation and the fluctuations. Experiments with even larger $\Delta_{SAS}$ could reveal this type of transition.

In this paper, we have discussed the Josephson-like tunneling at small voltages, where the theory and experiments are in good quantitative agreement with each other. On the other hand, it is clear (see Figs.~\ref{spielman63} and \ref{spielman53}) that our theoretical predictions differ from the experimental results at larger voltages. This discrepancy could be caused by the higher order terms in the Hamiltonian [Eq.~(\ref{startingpoint})], a slightly incorrect form of the vortex field correlation function [Eq.~(\ref{disordercor})], the effects caused by the dissipation (see Appendix \ref{calculation2}) or  an incoherent tunneling occurring inside compressible puddles of the electron liquid \cite{Eastham09}. The last possibility is supported by the experimental observation that at significantly larger voltages the tunneling I-V characteristic starts to look more and more similar to the corresponding I-V characteristic in the incoherent phase \cite{Eisenstein03, Spielman-diss}. Theoretical treatment of the tunneling at large voltages goes beyond the scope of the present paper.

\acknowledgments{We are indebted to A.~Stern for  important discussions and for  guiding us through the  literature on quantum Hall bilayers. We would like to acknowledge valuable  discussions with  W.~Dietsche, J.P.~Eisenstein,  X.~Huang, K.~von Klitzing, J.~Smet, E.~Thuneberg, Y.~Yoon and  D.~Zhang, and thank T.~Wright for a critical reading of the manuscript.
We thank J.P.~Eisenstein and I.B.~Spielman for
 providing the data used in figures \ref{spielman63}, \ref{spielman66}, \ref{spielman65}, \ref{spielman53}, \ref{theoT2}, and W.~Dietsche, L.~Tiemann, Y.~Yoon for
providing the experimental data used in figures \ref{stuttgarttemp} and \ref{counterflowcurtheo}.
This work was supported by the
BMBF (German Ministry of Education and Research) Grant No. 01BM0900. B.R. would like to acknowledge the hospitality of the Aspen Center for Physics, where part of this work was done.

\appendix

\section{Calculation of the I-V characteristics \label{calculation1}}

Similarly as in Josephson junctions \cite{Ingold-Nazarov} we can use Fermi's golden rule to calculate a forward tunneling rate
\begin{eqnarray}
 \overrightarrow{\Gamma}(V, B)&=&\frac{\lambda^2}{\hbar^2} \int_{-\infty}^\infty dt \int d^2 r_1 \int d^2 r_2 \ e^{ieVt/\hbar} e^{iQ_B (x_1-x_2)} \nonumber\\ && \hspace{-0.5 cm} \times \langle e^{i\varphi(\vec{r}_1, t)} e^{i\varphi_m(\vec{r}_1, t)} e^{-i\varphi(\vec{r}_2, 0)} e^{-i\varphi_m(\vec{r}_2, 0)} \rangle.
\end{eqnarray}
Due to the symmetry of the problem the backward tunneling rate is related to the forward tunneling rate as
\begin{equation}
\overleftarrow{\Gamma}(V, B)=\overrightarrow{\Gamma}(-V, -B).
\end{equation}
By assuming that $\varphi$ and $\varphi_m$ commute, we get
\begin{eqnarray}
I&=&e[\overrightarrow{\Gamma}(V, B)-\overleftarrow{\Gamma}(V, B)]\nonumber\\&=&\frac{2 \pi e \lambda^2 L^2}{\hbar}[S(Q_B, eV)-S(-Q_B, -eV)], \label{IVee}
\end{eqnarray}
where
\begin{eqnarray}
S(Q_B, eV)&=&\frac{1}{2\pi\hbar} \int_{-\infty}^\infty dt \int d^2 r \ e^{ieVt/\hbar} e^{iQ_B x} \nonumber\\&& \times \langle e^{i\varphi(\vec{r}, t)} e^{-i\varphi(\vec{0}, 0)} \rangle G_m(r,t). \label{S}
\end{eqnarray}
and
\begin{equation}
G_m(r,t)=\langle e^{i \varphi_m(\vec{r}, t)} e^{-i \varphi_m(\vec{0}, 0)} \rangle.
\end{equation}

By defining the Fourier components of the operator fields $\varphi$ and $\pi$ as
\begin{eqnarray}
\varphi_{\vec{k}}&=&\frac{1}{L} \int d^2 r e^{-i \vec{k} \cdot \vec{r}} \varphi(\vec{r},t)\\
\pi_{\vec{k}}&=&\frac{1}{L} \int d^2 r e^{i \vec{k} \cdot \vec{r}} \pi(\vec{r},t),
\end{eqnarray}
the commutation relations are $[\pi_{\vec{k}},\varphi_{\vec{k}'}]=-i\hbar \delta_{\vec{k}, \vec{k}'}$ and
the Hamiltonian defined by Eq.~(\ref{startingpoint}) can be rewritten as
\begin{eqnarray}
H&=& \int d^2 r \mathcal{H}\nonumber\\
&=&\frac{\pi_{0}^2}{2 M}+ \sum_{\vec{k}\ne 0} \bigg[ \frac{\pi_{\vec{k}} \pi_{-\vec{k}}}{2 M}+\frac{M \omega_{\vec{k}}^2}{2} \varphi_{\vec{k}} \varphi_{-\vec{k}}\bigg],
\end{eqnarray}
where $M=\hbar^2 \Gamma/e^2$ and $\omega_{\vec{k}}$ is the dispersion relation for the collective Goldstone mode given by Eqs.~(\ref{dispersion}) and (\ref{velocity}).

The creation and annihilation operators $a_{\vec{k}}$ and $a^\dag_{\vec{k}}$ are defined as
\begin{eqnarray}
a_{\vec{k}}&=&\sqrt{\frac{M \omega_{\vec{k}}}{2 \hbar}} \bigg(\varphi_{\vec{k}}+\frac{i}{M\omega_{\vec{k}}}\pi_{-\vec{k}}\bigg) \nonumber\\
a^\dag_{\vec{k}}&=&\sqrt{\frac{M \omega_{\vec{k}}}{2 \hbar}} \bigg(\varphi_{-\vec{k}}-\frac{i}{M\omega_{\vec{k}}}\pi_{\vec{k}}\bigg). \label{creaan}
\end{eqnarray}
It is easy to show that they satisfy commutation relations $[a_{\vec{k}}, a^\dag_{\vec{k}'}]=\delta_{\vec{k}, \vec{k}'}$, $[a^\dag_{\vec{k}}, a^\dag_{\vec{k}'}]=0$, $[a_{\vec{k}}, a_{\vec{k}'}]=0$ and allow to rewrite the Hamiltonian as
\begin{equation}
H=\frac{\pi_0^2}{2M}+\sum_{\vec{k}\ne 0} \hbar \omega_{\vec{k}} \bigg(a^\dag_{\vec{k}} a_{\vec{k}}+1/2 \bigg). \label{free+harmoscs}
\end{equation}

We can now express $\varphi(\vec{r}, t)$ as a Fourier series
\begin{equation}
\varphi(\vec{r}, t)=\frac{1}{L}\varphi_0(t)+\frac{1}{L}\sum \bigg(\varphi_{\vec{k}}(t)e^{i \vec{k} \cdot \vec{r}}+\varphi_{-\vec{k}}(t)e^{-i \vec{k} \cdot \vec{r}} \bigg), \label{summa}
\end{equation}
where the summation is taken so that each pair $\pm \vec{k}$ is counted only once. Thus the correlation function can be written as
\begin{eqnarray}
\langle e^{i\varphi(\vec{r}, t)} e^{-i\varphi(0, 0)} \rangle&=&\langle e^{\frac{i}{L}\varphi_0(t)}e^{-\frac{i}{L}\varphi_0(0)} \rangle \nonumber\\ && \hspace{-2.2 cm} \times \prod \langle e^{\frac{i}{L}[\varphi_{\vec{k}}(t)e^{i \vec{k} \cdot \vec{r}}+\varphi_{-\vec{k}}(t)e^{-i \vec{k} \cdot \vec{r}}]}e^{-\frac{i}{L}[\varphi_{\vec{k}}(0)+\varphi_{-\vec{k}}(0)]} \rangle, \nonumber\\ \label{corrapu}
\end{eqnarray}
where the product is taken similarly over the pairs $\pm \vec{k}$ to avoid double counting.

By using the Baker-Hausdorff relation, we can rewrite Eq.~(\ref{corrapu}) as
\begin{eqnarray}
\langle e^{i\varphi(\vec{r}, t)} e^{-i\varphi(0, 0)} \rangle&=&e^{-iC(\vec{r}, t)/2}\langle e^{\frac{i}{L} [\varphi_0(t)-\varphi_0(0)]} \rangle \nonumber\\ && \hspace{-2 cm} \times \prod \langle e^{\frac{i}{L}[\varphi_{\vec{k}}(t)e^{i \vec{k} \cdot \vec{r}}+\varphi_{-\vec{k}}(t)e^{-i \vec{k} \cdot \vec{r}}-\varphi_{\vec{k}}(0)-\varphi_{-\vec{k}}(0)]} \rangle, \nonumber\\ \label{corrapu2}
\end{eqnarray}
where
\begin{eqnarray}
C(\vec{r},t)&=&\frac{i}{L^2} \bigg\{ \big[\varphi_0(t), \varphi_0(0)\big]\nonumber\\&&\hspace{-1 cm}+\sum \big[\varphi_{\vec{k}}(t)e^{i \vec{k} \cdot \vec{r}}+\varphi_{-\vec{k}}(t)e^{-i \vec{k} \cdot \vec{r}}, \varphi_{\vec{k}}(0)+\varphi_{-\vec{k}}(0)\big] \bigg\}.\nonumber\\ \label{C}
\end{eqnarray}

By using the Wick theorem for equilibrium correlation functions, it is easy to show that an operator $\psi$, which is a linear combination of creation and annihilation operators, satisfies
\begin{equation}
\frac{d}{d\alpha} \langle e^{i \alpha \psi} \rangle=-\alpha \langle \psi^2 \rangle \langle e^{i \alpha \psi} \rangle,
\end{equation}
and therefore $\langle e^{i \psi} \rangle=e^{-\frac{1}{2}\langle \psi^2 \rangle}$.

By applying this result to Eq.~(\ref{corrapu2}), we can rewrite the correlation function as
\begin{equation}
\langle e^{i\varphi(\vec{r}, t)} e^{-i\varphi(0, 0)} \rangle=e^{-iC(\vec{r}, t)/2}  e^{-D(\vec{r},t)/2}, \label{corrapu4}
\end{equation}
where
\begin{eqnarray}
D(\vec{r},t)&=&\frac{t^2}{\beta M L^2}\nonumber +\frac{1}{L^2}\bigg\{ \sum \langle [\varphi_{\vec{k}}(t)e^{i \vec{k} \cdot \vec{r}} +\varphi_{-\vec{k}}(t)e^{-i \vec{k} \cdot \vec{r}}\\&& \hspace{0.5 cm}-\varphi_{\vec{k}}(0)-\varphi_{-\vec{k}}(0)]^2 \rangle \bigg\}.  \label{D}
\end{eqnarray}

The functions $C$ and $D$ can now be calculated by expressing $\varphi_{\vec{k}}$ in terms of creation and annihilation operators. After a straightforward calculation, we obtain
\begin{eqnarray}
C(\vec{r},t)&=&\frac{1}{L^2} \bigg\{ \frac{e^2 t}{\hbar \Gamma}+\sum \frac{ 2  u \hbar}{ \rho_s k} \sin(\omega_{\vec{k}}t)\cos(\vec{k} \cdot \vec{r}) \bigg\}\nonumber\\&=&\frac{1}{L^2} \bigg\{ \frac{e^2 t}{\hbar \Gamma}+\sum_{\vec{k}\ne 0} \frac{u \hbar}{ \rho_s k} \sin(\omega_{\vec{k}}t)\cos(\vec{k} \cdot \vec{r}) \bigg\} \nonumber\\&=&\frac{\hbar u}{L^2 \rho_s}\sum_{\vec{k}} \frac{1}{ k} \sin(\omega_{\vec{k}}t)\cos(\vec{k} \cdot \vec{r}). \label{Cexp}
\end{eqnarray}
Note that in each step we have changed the summation convention, so that finally the summation is over all $\vec{k}$ vectors. In the final form the term corresponding to $\vec{k}=0$ should be understood as a limit $k \to 0$ for the given expression.

It is easy to show that $C$ is independent on the direction in the x,y-plane $C(\vec{r}, t)=C(r, t)$ and that $C(r,-t)=-C(r,t)$. Moreover, for $t>0$
\begin{equation}
C(r, t)=\frac{\hbar}{2\pi \rho_s} \theta(ut-r)\frac{1}{\sqrt{t^2-\big(\frac{r}{u}\big)^2}}. \label{apulisC}
\end{equation}
The last expression is easiest to prove by calculating the Fourier expansion for the right side of Eq.~(\ref{apulisC}).

Similarly, we get
\begin{equation}
D(\vec{r},t)=\sum_{\vec{k}}  \frac{\hbar u}{L^2 \rho_s k}[1-\cos(\omega_{\vec{k}}t) \cos(\vec{k} \cdot \vec{r})] \coth(\beta \hbar \omega_{\vec{k}}/2), \label{Dexp}
\end{equation}
where the $\vec{k}=0$ term should be calculated by setting $\cos(\vec{k} \cdot \vec{r}) \to 1$ and taking the limit $k \to 0$ for the rest of the expression. It is again easy to show that $D(\vec{r},t)=D(r,t)$ and $D(r,-t)=D(r,t)$.

By using Eqs.~(\ref{IVee}), (\ref{S}) and (\ref{corrapu4}) and the properties $C(\vec{r}, t)=C(r, t)$, $D(\vec{r},t)=D(r,t)$, $C(r, -t)=-C(r, t)$, $D(r, -t)=D(r,t)$ and $G_m(r,-t)=G_m(r,t)$, we get equations (\ref{main1}), (\ref{main2}) and (\ref{main3}).

\section{Derivation of the I-V characteristics using Sine-Gordon equation \label{calculation2}}

The Lagrangian density corresponding to Hamiltonian density (\ref{startingpoint}) is
\begin{equation}
\mathcal{L}=\frac{\hbar^2 \Gamma}{2 e^2} \dot{\varphi}^2-\frac{\rho_s}{2}(\nabla \varphi)^2+2 \lambda \cos \bigg[\varphi+\varphi_m+Q_B x+\frac{eVt}{\hbar} \bigg].
\end{equation}
The equation of motion for $\varphi$ can be written as a Sine-Gordon equation
\begin{equation}
\nabla^2 \varphi-\frac{1}{u^2}\partial_t^2 \varphi-\hat{\gamma} \varphi=\frac{1}{\lambda_J^2} \sin \bigg[\varphi+\varphi_m+Q_B x+eVt/\hbar \bigg], \label{dissipativeclass}
\end{equation}
where we have introduced an operator $\hat{\gamma}$, which describes the dissipative terms in our model. The actual mechanism of the dissipation is not well-understood, but several different dissipative terms have been proposed \cite{Fogler-Wilczek01, Su10, Fil09}. Here we keep the dissipative terms unspecified for as long as possible.

The idea is to formulate the perturbative tunneling theory so that we would only need to assume that the variations of $\varphi$ are small within certain reasonably large space and time scales.
More concretely, we divide the space-time into domains (labeled with index $i$) of size $T_i$ (time-interval) and $L_i^2$ (spatial area). We denote the average value of $\varphi$ inside the domains as $\overline{\varphi}_i$, and assume that the variations within the domain are small, $\delta \varphi_i(\vec{r}_i, t_i)=\varphi(\vec{r}_i, t_i)-\overline{\varphi}_i \ll 1$. The variations $\delta \varphi_i$ inside each domain $i$ are not entirely independent on the other domains because the domains can be dynamically coupled with each other.
Here we assume that the domains are large enough that the boundary conditions are not important. Although the boundary conditions strongly affect the state of the whole system in the case of static vortex field \cite{Eastham10}, we believe that our approximation can be justified in the typical experimental situation where the fluctuations of the vortex field are important.

Inside each space-time domain we need to solve an equation
\begin{equation}
\nabla^2 \delta \varphi_i-\frac{\partial_t^2 \delta \varphi_i}{u^2}-\hat{\gamma} \delta \varphi_i
=\frac{1}{\lambda_J^2} \sin \bigg[\overline{\varphi}_i+\varphi_m+Q_B x+\frac{eVt}{\hbar} \bigg]. \label{perturbative}
\end{equation}
The time average of tunneling current $I(V)$ can then be calculated from equation
\begin{eqnarray}
\langle I(V) \rangle&=&\frac{e}{\hbar} \frac{\Delta_{SAS}}{4 \pi l_B^2}  \sum_i \frac{1}{T_i} \int dt_i \int d^2 r_i \nonumber\\&& \hspace{-1.6 cm} \times \langle \sin \big[\delta \varphi_i(\vec{r}_i, t_i)+\overline{\varphi}_i+\varphi_m(\vec{r}_i, t_i)+Q_B x_i+eVt_i/\hbar \big] \rangle. \nonumber\\ \label{defVIapp}
\end{eqnarray}
Here $\sum_i T_i \to \infty$ is the total time, $\langle...\rangle$ denotes the ensemble average over different realizations of the vortex field and the integrations are over the space-time domains.

It is easy to show that a particular solution of Eq.~(\ref{perturbative}) can be written as
\begin{eqnarray}
\delta \varphi_i(\vec{r}_i, t_i)&=&C_i+\frac{1}{\lambda_J^2} \int \int \ d^2 r' \ dt'  \ G(\vec{r}_i-\vec{r}', t_i-t') \nonumber\\ && \hspace{-0.8 cm} \times \sin \bigg[\overline{\varphi}_i+\varphi_m(\vec{r}', t')+Q_B x'+eVt'/\hbar \bigg], \label{parsolapp}
\end{eqnarray}
where
\begin{eqnarray}
G(\vec{r}, t)&=& \frac{1}{L^2} \frac{1}{2 \pi} \sum_{\vec{k}}  \int d \omega \ G(\vec{k}, \omega) e^{i\vec{k} \cdot \vec{r}} e^{-i \omega t},  \nonumber\\
 G(\vec{k}, \omega)&=&\frac{1}{-k^2 +\frac{\omega^2}{u^2} +i\gamma(\omega, \vec{k})}
\end{eqnarray}
and
\begin{eqnarray}
C_i&=&-\frac{1}{T_i L_i^2}\int dt_i \int d^2 r_i \  \frac{1}{\lambda_J^2} \int \int \ d^2 r' \ dt'  \  \nonumber\\ && \hspace{-1 cm} \times G(\vec{r}_i-\vec{r}', t_i-t') \sin \bigg[\overline{\varphi}_i+\varphi_m(\vec{r}', t')+Q_B x'+eVt'/\hbar \bigg] \nonumber\\
\end{eqnarray}
is a constant guaranteeing that $\overline{\varphi}_i$ is the average value of $\varphi$ within the domain $i$ i.e.
\begin{equation}
\int dt_i \int d^2 r_i \delta \varphi_i(\vec{r}_i, t_i)=0.
\end{equation}
We have also assumed that the dissipation operator acting on the plane waves satisfied the following equation
\begin{equation}
\hat{\gamma} e^{i\vec{k} \cdot \vec{r}} e^{-i \omega t}=-i\gamma(\omega, \vec{k}) e^{i\vec{k} \cdot \vec{r}} e^{-i \omega t}.
\end{equation}
We assume that the dissipation is isotropic in the ($x,y$)-plane, $\gamma(\omega, \vec{k})=\gamma(\omega,k)$. Moreover, we require that $G(-\omega, -\vec{k})= G^*(\omega, \vec{k})$, which gives $\gamma(-\omega, k)=-\gamma(\omega,k)$.

By assuming that the instantaneous value of the vortex field $\varphi_m(\vec{r},t)$ and the average quantities $C_i$ and $\overline{\varphi}_i$ are statistically independent, we obtain using
Eqs.~(\ref{defVIapp}) and (\ref{parsolapp}) that the tunneling current $I(V)$ can be written as
\begin{eqnarray}
\langle I(V) \rangle
&=&\frac{1}{2}\frac{e}{\hbar} \frac{\Delta_{SAS}}{4 \pi l_B^2} \frac{1}{\lambda_J^2}  \textrm{Im} \bigg\{ \nonumber\\ && \hspace{-1.5 cm} \sum_i \frac{1}{T_i} \int dt_i \int d^2 r_i  \ \int \int \ d^2 r' \ dt'  \ G(\vec{r}_i-\vec{r}', t_i-t') \nonumber\\ && \hspace{-1.5 cm} \times \bigg\langle e^{i[\varphi_m(\vec{r}', t')+Q_B x'+eVt'/\hbar -\varphi_m(\vec{r}_i, t)-Q_B x_i-eVt_i/\hbar ]} \bigg\rangle \bigg\}.  \nonumber\\ \label{VIapu}
\end{eqnarray}
Here we have taken into account that the vortex field tends to stay constant over some time and space interval and therefore the differences of the vortex field values $\varphi_m(\vec{r}', t')-\varphi_m(\vec{r}_i, t)$ are correlated. On the other hand, the actual value of the vortex field $\varphi_m(\vec{r}, t)$ can be arbitrary, and therefore only the terms proportional to $e^{\pm i[\varphi_m(\vec{r}', t')-\varphi_m(\vec{r}_i, t)]}$ can contribute to the average current. Physically this means that the fluctuations of the vortex field destroy the supercurrent.

By assuming also that correlation function for the vortex field is given by Eq.~(\ref{disordercor}), we obtain [$R=r/\xi$, $q=k \xi$, $\Omega=\omega \xi/u$, $\gamma'(\Omega, q)=\gamma(\Omega u/\xi, q/\xi)$]
\begin{eqnarray}
\langle I \rangle
&=&I_0' \int_0^\infty d q \  \int_0^\infty d \Omega \ \bigg\{ \frac{\gamma'(\Omega, q) \xi^2}{(\Omega^2-q^2 )^2 +[\gamma'(\Omega, q)\xi^2]^2} \nonumber\\ &&  \times \frac{2 q}{\pi} \bigg[\frac{\alpha}{\alpha^2+(-\Omega+V/V_0 )^2}-\frac{\alpha}{\alpha^2+(\Omega+V/V_0)^2} \bigg] \nonumber \\ && \times  \int  d R \ R e^{-R} J_0(Q_B \xi R) J_0(qR) \bigg\}, \label{dissVI}
\end{eqnarray}
where  $V_0$ and $\alpha$ are given by Eqs.~(\ref{V0}) and (\ref{alpha}), respectively. The current scale is now determined by
\begin{equation}
I_0'=\frac{e}{\hbar} \frac{\Delta_{SAS}}{16 \pi l_B^2} \frac{L^2 \xi^2}{\lambda_J^2}=\frac{e}{\hbar} \frac{\xi^2 L^2}{l_B^4} \frac{\Delta_{SAS}^2}{\rho_s}\frac{1}{64 \pi^2}, \label{I0'}
\end{equation}
which differs from the expression (\ref{I0exp}) by a factor $e^{-D_0/2}$.

If we further assume that $\gamma'(\Omega, q) \xi^2 \ll 1$, we can use
\begin{equation}
\lim_{\epsilon \to 0} \frac{1}{\pi} \frac{\epsilon}{(y^2-y_0^2)^2+\epsilon^2}=\frac{1}{2|y_0|}[\delta(y+y_0)+\delta(y-y_0)]
\end{equation}
to obtain
\begin{eqnarray}
\langle I\rangle
&=&I_0' \int d q \  \bigg[\frac{\alpha}{\alpha^2+(V/V_0-q)^2}-\frac{\alpha}{\alpha^2+(V/V_0+q)^2} \bigg] \nonumber\\ && \times \int  d R  \ R e^{-R} J_0(Q_B \xi R) J_0(q R).
\end{eqnarray}
The only difference to the expression (\ref{altapp}) is that the current scale $I_0$ determined by Eq.~(\ref{I0exp}) is now replaced by $I_0'$ given by Eq.~(\ref{I0'}). We see from Eq.~(\ref{dissVI}) that in the general case also the dependence of the dissipation on the frequency and momentum of the pseudospin waves can affect the I-V characteristic.

In order to determine the limits of validity of the perturbation theory we need to calculate $\langle [\varphi(\vec{r}, t)-\varphi(\vec{0}, 0)]^2\rangle=\langle [\delta \varphi(\vec{r}, t)-\delta \varphi(\vec{0}, 0)]^2\rangle=2\langle \delta \varphi^2(\vec{r}, t) \rangle-2 \langle \delta \varphi(\vec{r}, t) \delta \varphi(\vec{0}, 0) \rangle$.
Using similar methods as above, we obtain that the perturbation theory is controlled by a condition
\begin{equation}
\frac{\xi^4}{\lambda_J^4} \frac{1}{\gamma \xi^2} \ll 1, \label{parameter}
\end{equation}
where for simplicity we have assumed that the dissipation $\gamma$ does not depend on frequency and momentum.


\begin{thebibliography}{99}
\bibitem{Fitdetail}
We vary $I_0$ and determine $\alpha$ for each temperature as the average of the two values obtained by fitting the height and width of the conductance peak. We choose the value for $I_0$ by minimizing a cost function $\sum [|\log(G_0(\alpha_i)/G_i)/\log(G_{\max}/G_{\min})|+|(V_{HWHM}(\alpha_i)-V_i)/V_0|$, where the summation is over the five lowest temperatures, $\alpha_i$ are the values of $\alpha$ fitted at each temperature, $G_0(\alpha_i)$ and $G_i$ are the theoretically calculated and experimentally measured peak conductances, $G_{\max}$ and $G_{\min}$ are the experimentally measured maximum and minimum peak conductances in the temperature interval $T=25-300$ mK, and $V_{HWHM}(\alpha_i)$ and $V_i$ are the theoretically calculated and experimentally measured half widths at half maximums for the conductance peaks.
\bibitem{comment}
There are some sample to sample variations also in other parameters and therefore it is not meaningful to try to make the prediction for $I_0$ more accurate. However, we point out that it would be very useful to study experimentally the scaling of $I_0$ as a function of $\Delta_{SAS}$ by using samples which have otherwise as identical parameters as possible.
\end{thebibliography}
\end{document}